%% file: Brand_AA_2020_39506_final.tex
\newcommand{\TAStar}{$T^*\hskip-2pt_{\rm A}$}

\newcommand{\pas}{.\hskip-2pt$^{\prime\prime}$}

\newcommand{\gsim}{\;\lower.6ex\hbox{$\sim$}\kern-7.75pt\raise.65ex\hbox{$>$}\;}
\newcommand{\lsim}{\;\lower.6ex\hbox{$\sim$}\kern-7.75pt\raise.65ex\hbox{$<$}\;}

\documentclass{aa}  
\usepackage{graphicx}
\usepackage{txfonts}
\usepackage[english]{babel}  
\usepackage{graphicx}  
\usepackage[utf8]{inputenc}  
\usepackage{microtype}  
\usepackage{booktabs}  
\usepackage{rotating}  
\usepackage{lscape}
\usepackage[squaren]{SIunits}
\usepackage{natbib}
\usepackage{color}
\usepackage{amsmath}
\usepackage{amssymb}

\input{new_commands_aa.tex}

\begin{document}

   \title{A possible far-ultraviolet flux-dependent core mass function in \object{NGC~6357}
   \thanks{Table~\ref{tab:686cores} is only available in electronic form at the CDS via anonymous ftp to cdsarc.u-strasbg.fr(130.79.128.5) or via http://cdsarc.u-strasbg.fr/viz-bin/qcat?J/A+A/.....}}


   \author{J. Brand\inst{1} 
          \and
          A. Giannetti\inst{1}\fnmsep\thanks{Present address: Energy Way s.r.l., Modena, Italy; andrea.giannetti@energyway.it} 
          \and          
          F. Massi\inst{2} 
          \and
          J.G.A. Wouterloot\inst{3}
          \and
          C. Verdirame\inst{4}
          }

   \institute{INAF - Istituto di Radioastronomia \& Italian ALMA Regional Centre, via P. Gobetti 101, 40129 Bologna, Italy\\
              \email{brand@ira.inaf.it}
         \and INAF - Osservatorio Astrofisico di Arcetri, Largo E. Fermi 5, 50125 Firenze, Italy
         \and East Asian Observatory,
         660 N. A'ohoku Place, Hilo, Hawaii 96720, USA
         \and Dip.to di Fisica e Astronomia, Universit\`a di Bologna, via P. Gobetti 93/2, 40129 Bologna, Italy
             }

   \date{Received date; accepted date}

 
  \abstract
   {NGC6357 is a galactic star-forming complex ($d \sim 1.7$ kpc) composed of several H\textsc{ii} regions, a few young stellar clusters, and giant molecular clouds. In particular, the H\textsc{ii} regions \object{G353.2+0.9}, \object{G353.1+0.6}, and \object{G353.2+0.7} are associated with three young clusters; the most prominent of these, \object{Pismis 24}, contains some of the most massive stars known.}
   {We aim to derive the properties of the densest compact gas structures (cores) in the region as well as the effects of an intense far-ultraviolet (FUV) radiation field on their global properties.}
   {We mapped the NGC6357 region at 450 and 850 $\mu$m with SCUBA-2 and in the CO(3-2) line with HARP at the JCMT. We also made use of the Herschel Hi-GAL data at 70 and 160 $\mu$m. We used the algorithm Gaussclumps to retrieve the compact cores embedded in the diffuse sub-millimetre emission and constructed their spectral energy distribution from 70 to 850 $\mu$m, from which we derived mass and temperature. We divided the observed area into an 'active' region (i.e. the eastern half, which is exposed to the FUV radiation from the more massive members of the three clusters) and a 'quiescent' region (i.e. the western half, which is less affected by FUV radiation). We compared the core mass functions and the temperature distributions in the two areas to look for any differences that could be due to the different levels of FUV radiation.}
   {We retrieved 686 dense cores, 411 in the active region and 275 in the quiescent region, with an estimated mass completeness limit of $\sim 5$ M$_{\sun}$. We also attempted to select a sample of pre-stellar cores based on cross-correlation with 70 $\mu$m emission and red WISE point sources, which unfortunately is biased due to distance, emission at 70 $\mu$m from the dust on the surface of the cores that is heated by the FUV radiation, and saturation in the WISE bands. Most of the cores above the mass completeness limit are likely to be gravitationally bound. The fraction of gas in dense cores is very low, $1.4$ \%. We found a mass-size relation $\log(M/M_{\sun}) \sim a \times \log (D/{\rm arcsec})$, with $a$ in the range $2.0$-$2.4$, depending on the precise selection of the sample. 
   The temperature distributions in the two sub-regions are clearly different, peaking at $\sim 25$ K in the quiescent region and at $\sim 35$ K in the active region. The core mass functions are different as well, at a $2 \sigma$ level, consistent with a Salpeter initial mass function in the quiescent region and flatter than that in the active region. The dense cores lying close to the H\textsc{ii} regions are consistent with pre-existing cores being gradually engulfed by a photon dominated region and photoevaporating. A comparison of the obtained distribution of core masses with those derived from simulations of cloud-cloud collisions yields no conclusive evidence of ongoing cloud-cloud collisions.}
   {We attribute the different global properties of dense cores in the two sub-regions to the influence of the FUV radiation field.}


   \maketitle
%
\noindent

\section{Introduction}
The ionising radiation of early type (massive) stars generates a photon dominated region (PDR) where it hits a nearby molecular cloud. Molecular clouds are complex entities, having a self-similar structure over a wide range of scales (e.g. \citealt{2007msl..confE..95O}).
With increasing  spatial resolution molecular clouds appear fragmented in substructures of smaller angular dimensions. Hence, the molecular material is typically very clumpy, implying that a large part of the molecular cloud can be exposed to far-ultraviolet (FUV) radiation ($6 < h\nu < 13.6$ eV), which can penetrate much deeper than if the cloud were homogeneous, making the PDR volume vastly larger. The radiative and mechanical (winds) action of massive stars can evaporate existing structures, but it may also trigger the formation of a new generation of stars in the surrounding material, for example through radiation-driven implosion (RDI; see e.g. \citealt{2011ApJ...736..142B}).

The importance of gas fragmentation in star-forming regions has been reinforced by the discovery that the smallest ($\sim 0.1$ pc) and densest structures ($\ga 10^{4} - 10^{5}$ cm$^{-3}$), named cores, inside giant molecular clouds follow a well-defined mass distribution (core mass function: CMF) resembling the stellar initial mass function (IMF; see
\citealt{2009ApJ...699..742R}; \citealt{2009ApJ...705L..95I}; \citealt{Andre+14_PPVI}
and references therein). We note that cores are embedded in molecular clumps, which are larger
($0.3 - 3$ pc) and less dense ($10^{3} - 10^{4}$ cm$^{-3}$) gas concentrations
(see e.g. \citealt{2007ARA&A..45..339B}). 
Although the link between the CMF and the IMF is still debated, there seems to be growing consensus that the IMF derives from the CMF through further fragmentation and star formation (taking into account the formation of binaries and multiple systems; \citealt{2008A&A...477..823G}), provided that the star-forming efficiency (SFE) of each core is less than $\sim 30-50$ \% (\citealt{2007A&A...462L..17A}).
On the other hand, \cite{2018NatAs...2..478M} have recently found that the CMF in the mini-starburst galactic region W43-MM1 challenges this scenario of direct proportionality, in that its slope is "markedly shallower" than that of the IMF, for core masses greater than about a solar mass; variations in the CMF are indeed possible in different environments.

If the CMF does in fact evolve into the stellar IMF, then the distribution of stellar masses will be linked to the origin of dense cores, and different CMFs may significantly affect the mass distribution of the final stellar population. 
The characteristics of the CMF (its slopes as well as the presence of a turn-over and its location in mass) are likely to depend on conditions in the molecular clouds in which the cores are embedded, the properties of which, in turn, depend on the environment in which they were formed. Therefore, it has to be expected that if intense UV radiation has any effect on star formation, it must be by primarily affecting the CMF in the exposed molecular cloud(s).

The region of interaction between young massive stars and a molecular cloud is thus of great interest. Hydrodynamical simulations (e.g. \citealt{2007MNRAS.377..535D};  \citealt{2011ApJ...736..142B}; \citealt{2011MNRAS.414.1747A}; \citealt{2020ApJ...904..192W}; \citealt{2020MNRAS.499..668G})
have shown that star formation may be either triggered or halted in clouds exposed to intense FUV radiation. The FUV flux intensity is possibly a key element in determining one of the two scenarios. Feedback from young massive stars is believed to be instrumental in regulating the SFE in galaxies (\citealt{2014PhR...539...49K}; \citealt{2015A&A...573A.112M}). \cite{2017MNRAS.466.5011L} propose that the removal of molecular cloud envelopes by stellar feedback can significantly decrease a cloud's SFE. Thermal feedback affects the way in which a molecular cloud fragments and can lead to a top-heavy IMF (\citealt{2014PhR...539...49K}). 
A way of observationally assessing the effects of intense FUV radiation fields on molecular gas is therefore to determine whether the CMF near to a PDR significantly differs from that observed in more quiescent regions (i.e. in regions less exposed to intense stellar FUV radiation).

NGC6357 is a galactic complex of molecular clouds and H\textsc{ii} regions 
that hosts several active star-forming sites. \cite{Massi+15_aap573_95} discussed the issue of distance, concluding that NGC6357 is located 1.7~kpc away. A more reliable assessment can now be based on GAIA DR2 data (\citealt{2018A&A...616A...1G}). \cite{2018A&A...616L..15X} retrieved the parallaxes of a large sample of OB stars from the GAIA DR2 catalogue. Those associated with NGC6357 lie in the range $1.7 - 2.0$ kpc, with quoted uncertainties of $\sim 10$ \%. In particular,
\cite{2020A&A...633A.155R} derive a distance of $1.77 \pm 0.12$ kpc
and \cite{2020A&A...643A.138M} a distance of $1690^{+130}_{-110}$ pc,
both using GAIA data.
This definitely rules out a distance as large as $2.56$ kpc, which was derived by \cite{2001AJ....121.1050M}.
Thus, we will safely continue using $d = 1.7$ kpc as in \cite{Massi+15_aap573_95}, which is likely to be correct within $\sim 10$ \%. 

Figure~1 in \cite{2012A&A...538A..41G}
shows that the molecular gas in NGC6357 is arranged in a ring surrounding a large ($\sim$ 15~pc $\times$ 10~pc) cavity (or a collection of connected smaller cavities), possibly shaped by feedback from one or more star clusters (\citealt{2007ApJS..168..100W}).
In the northern part of this complex is located G353.2+0.9 (cf. Fig.~3 in \citealt{Massi+15_aap573_95}), 
which is the most luminous region found in the ring at wavelengths from radio to optical. Just $55"$ (0.45~pc at a distance of 1.7~kpc) 
south of G353.2+0.9 lies the massive, young open star cluster Pismis~24, which contains some of the brightest and bluest stars known (O3.5; \citealt{Maiz-Apellaniz+07}).
G353.2+0.9 is a blister H\textsc{ii} region seen nearly face-on, with an extended ionisation front (IF) or PDR on the side facing the ionising stars and with an elephant trunk seen prominently against the ionised gas. South-east of G353.2+0.9 is G353.1+0.6, another H\textsc{ii} region with an IF or PDR, seen edge-on, associated with the cluster AH03J1725–34.4 (\citealt{Massi+97}; \citealt{Massi+15_aap573_95}).
East of these two is another H\textsc{ii} region, G353.2+0.7, associated with cluster B of \cite{Massi+15_aap573_95}. G353.2+0.9, G353.1+0.6, and G353.2+0.7 are ideal targets for studying the molecular gas fragmentation expected in FUV-exposed gas, considering both the extent of the PDR around the giant  molecular clouds observed and the relatively close distance.

Several tracers of molecular density and temperature were observed with the SEST (\citealt{Massi+97}; \citealt{2012A&A...538A..41G}) 
in a 3\arcmin\ $\times$ 3\arcmin\ area around G353.2+0.9 (total mass $\sim 2000~\msun$). An analysis of the complex molecular emission profiles led to the identification of at least 14 clumps in the area mapped, indicating that the gas is indeed fragmented. 
However, the angular resolution (21--54$^{\prime\prime}$) of these single-dish observations is insufficient to resolve some clumps for which there are indications of smaller-scale structures.
Near-infrared ($J$, $H$, and $K$) images of G353.2+0.9 obtained with the ESO-NTT telescope (\citealt{Massi+15_aap573_95}),
indicate the presence of several embedded objects coinciding with the brightest part of the IF, which are younger than the stars of Pismis 24. This suggests that star formation in the region has been triggered by the action of the massive stars.
\cite{2012A&A...538A..41G} also used the ATLASGAL map of the region at 870 $\mu$m \citep{Schuller+09_aap504_415}, showing the most prominent clumps seen at millimetre lines. However, although its spatial resolution ($\sim 18^{\prime\prime}$, $\sim 0.2$~pc) is enough to resolve gas at the typical scale of young small embedded clusters, this map has a poor sensitivity (100~mJy/beam), and most of the clumps found at millimetre lines went undetected.
 
NGC6357/G353.2+0.9 represents a rather extreme environment, with 2-3 O3.5f stars and many later-type O stars at $\sim 1$~pc from the cloud. We can expect the intense FUV field to affect the nearby gas, for example by rapidly photoevaporating the smallest cores and thus producing a flatter CMF. Other parts of the complex are more quiescent and have no clear signs of star formation, and therefore no intense FUV-radiation field is expected there. Because of this contrast, NGC6357 may be one of the best locations to study the influence of FUV radiation on star formation. In this paper we identify dense cores from dust-continuum data in an area of about 0.5 square degrees in NGC6357. Combining these data with Herschel Hi-GAL \citep{Molinari+10} images we construct the spectral energy distribution (SED) for each core. Fitting a greybody to those SEDs yields  temperatures and masses for all cores, from which a mass function is derived. By dividing the field into a sub-field where a more intense FUV field is expected and a sub-field less exposed to FUV radiation, we carry out a comparison between the mass functions from the two sub-fields.

The layout of the paper is the following. In Sect.~\ref{sect:scuba2_obs} we describe our observations and data reduction. Our results are presented in Sect.~\ref{ou:res} and discussed in Sect.~\ref{dis:sec}. Finally our conclusions are listed in Sect.~\ref{sum:conc}. 

  \section{Observations and data reduction} 
  \label{sect:scuba2_obs}
\begin{figure}
\rotatebox{-90}{\includegraphics[width=6cm]{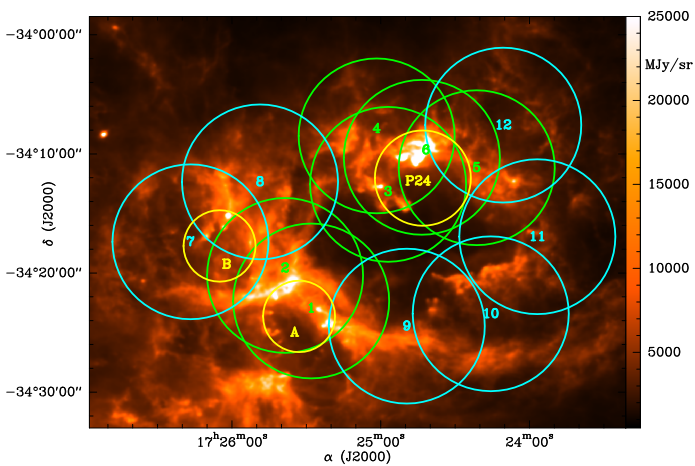}}
\caption{Herschel HiGAL $160~\mu$m map (colour scale) 
with fields observed in 2012 (1--6; green) and 2015 (7--12; blue). The location and approximate size of star clusters are marked in yellow \citep{Massi+15_aap573_95}. The linear size of the region shown is approximately 22 $\times$ 17~pc$^2$.}
\label{fig:daisyfields}
\end{figure}

We used the Submillimetre Common-User Bolometer Array 2 
(SCUBA-2: \citealt{2013MNRAS.430.2513H})
at the James Clerk Maxwell Telescope (JCMT) to observe 12 partially overlapping fields, simultaneously at 450~$\mu$m and 850~$\mu$m: on 13 July 2012 (fields 1--6; project m12an004) and on 6 and 23 April and 22 May 2015 (fields 7--12; project m15ai093). 
The observations were carried out using a so-called CV Daisy pattern\footnote{https://www.eaobservatory.org/jcmt/instrumentation/continuum/scuba-2/observing-modes/}; each daisy-field observation took 30~minutes. Each field has a radius of about 8~arcmin, and together they cover the entire ring-like structure of NGC6357, including three star clusters, as illustrated in Fig.~\ref{fig:daisyfields}.
The images were reduced using {\it SMURF} and associated packages contained in STARLINK\footnote{http://www.starlink.ac.uk}, following the Cookbook version SC/21.1\footnote{http://www.starlink.ac.uk/docs/sc21.htx/sc21.html} \citep{2013MNRAS.430.2545C}.
After reduction, the fields were mosaicked together. The resulting sensitivity varies with location, and except at field edges is $\lsim 4-15$~mJy/beam 
(at  850~$\mu$m), and 60-400 mJy/beam (at 450~$\mu$m). We used the flux conversion factors (FCFs; see \citealt{2013MNRAS.430.2534D})
listed in Table~\ref{fcf}. 
These were derived from measurements of CRL2688 taken close in time to each series of observations. 
The Starlink SDF files were then converted into FITS format for further analysis. The (effective) beam size of SCUBA-2 is 9\arcsec $\pm1$\arcsec\ at 450~$\mu$m and 14\arcsec $\pm1$\arcsec\ at 850~$\mu$m (i.e. $\sim 0.07$~pc and $\sim 0.12$~pc, respectively).
To complement these observations and construct SEDs, we used the $70~\mu$m and $160~\mu$m images from the Herschel/Hi-GAL survey \citep{Molinari+10}, with beam sizes of 5\arcsec\ and 11\pas5, respectively.

\begin{table}
 \caption[]{Adopted flux conversion factors (FCFs).}
 \label{fcf} 
\begin{flushleft}
\begin{tabular}{lccc}
 \hline
 \noalign{\smallskip}
Field & date & 850~$\mu$m & 450~$\mu$m \\
       &(ddmmyy) & \multicolumn{2}{c}{(Jy\,pW$^{-1}$\,beam$^{-1}$)} \\ 
\noalign{\smallskip}
\hline
\noalign{\smallskip}
1$-$6   & 130712 & 537 & 492 \\
10, 11  & 060415 & 592 & 627 \\
7, 8, 9 & 230415 & 544 & 581 \\
12      & 220515 & 525 & 460 \\
\noalign{\smallskip}
\hline
\end{tabular}
\end{flushleft}
\end{table}

\subsection{Correction for line-contamination \label{cocorr}}
The 850~$\mu$m emission is potentially contaminated by 
\CO(3--2) line emission at 345.8~GHz. The continuum band is very wide (ca. 39~GHz, see \citealt{2014SPIE.9153E..23N}) and lines have to be bright and broad in order to significantly affect the continuum emission. Several studies (e.g. \citealt{2013ApJ...767..126S})
show that the contamination is typically less than $\sim 70$~mJy/beam, but may reach $\sim 150$~mJy/beam ($10-30$ times the rms level in our map), and would thus affect especially the lower-level emission and hence the lower-mass cores, which are important to define the low-mass turnover point in the CMF. As there are numerous locations with star formation in NGC6357, outflow emission, and therefore contamination at relevant levels, is to be expected.  

\noindent
On the other hand, \CO(6--5) falls at the edge of the 450~$\mu$m band. No significant contamination is expected (also considering the high noise level of the 450~$\mu$m continuum data).

We used the HARP receiver at the JCMT (see \citealt{2009MNRAS.399.1026B})
to map the \CO(3--2) emission in the $\sim$1900\arcsec\ $\times$ 1700\arcsec\ region covered by the SCUBA-2 observations on 2, 7, and 20 May and 9 and 18 July 2014 (project m14au032). Observations were made in 'basket-weave' raster-scanning mode. HARP has 16 SIS receptors, 14 of which were operational at the time of the observations. The reference position 
(17:25:36.2 $-$33:55:17) was checked using frequency switching and showed only the telluric CO line; rms = 0.3~K (\TAStar) at 0.3~$\kms$ resolution. We used the ACSIS backend with a bandwidth of 1000~MHz, which provided 2048 channels with a spectral resolution of 0.488~MHz (0.423~$\kms$ at 345.796 GHz). The data were reduced using the ORAC-DR pipeline
\citep{2015MNRAS.453...73J};
the maps were then converted into CLASS format for further analysis. The main-beam efficiency of HARP is 0.64, and all line intensities were converted into a $\tmb$ scale. The average rms in the CO map is 2.0~K, on the $\tmb$ scale. The beam size of HARP at 345 GHz is 14\arcsec.

We followed the procedure outlined by \cite{2018ApJS..234...22P} to correct continuum
data for line contamination, using the HARP \CO(3-2) integrated intensity map as a negative fake map to be combined with the SCUBA-2 data in the map maker (see also \citealt{2012MNRAS.426...23D}).
For this we made use of a script kindly provided by Dr. Parsons. 

\section{Results}
\label{ou:res}
\begin{figure}
\includegraphics[angle=-90,width=\columnwidth]{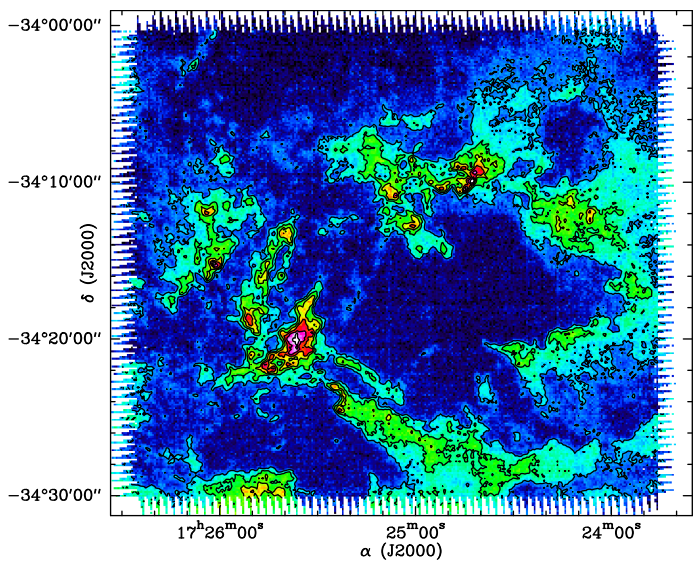}
\caption{$^{12}$CO(3--2) integrated emission ($\int \tmb$ dV; $-20$ to $+20$~$\kms$). Contours values are 80(80)470~K$\kms$ (low(step)high).}
\label{fig:CO32map}
\end{figure}

\begin{figure*}
\includegraphics[width=\textwidth]{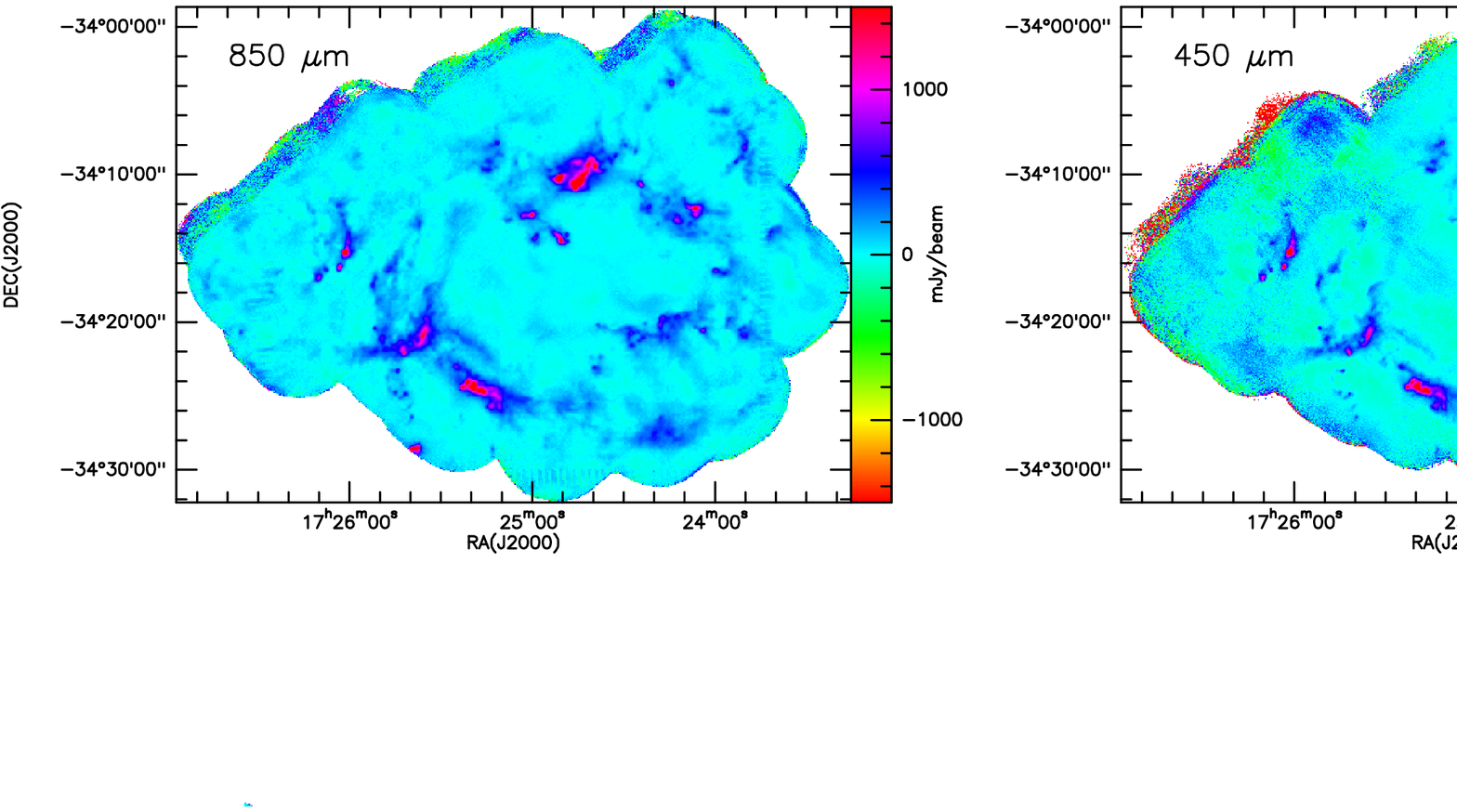}
\caption{SCUBA-2 observations of the dust emission. Left: $850~\mu$m mosaic; the emission has been corrected for the integrated $^{12}$CO(3--2) line emission. Right: $450~\mu$m mosaic. Units are mJy/beam.}
\label{fig:submm-maps}
\end{figure*}

The \CO(3--2) map used to correct the 850~$\mu$m continuum emission is shown in Fig.~\ref{fig:CO32map}. 
Following the procedure outlined in Sect.~\ref{cocorr}, we obtained CO-corrected 850~$\mu$m maps for each of the 12 observed fields. 
The individual fields at each wavelength were then combined into the single map shown in Fig.~\ref{fig:submm-maps} using the task 'wcsmosaic'. 

\subsection{Core identification}\label{sec:clump_id}
    	
\begin{figure}
\includegraphics[width=\columnwidth]{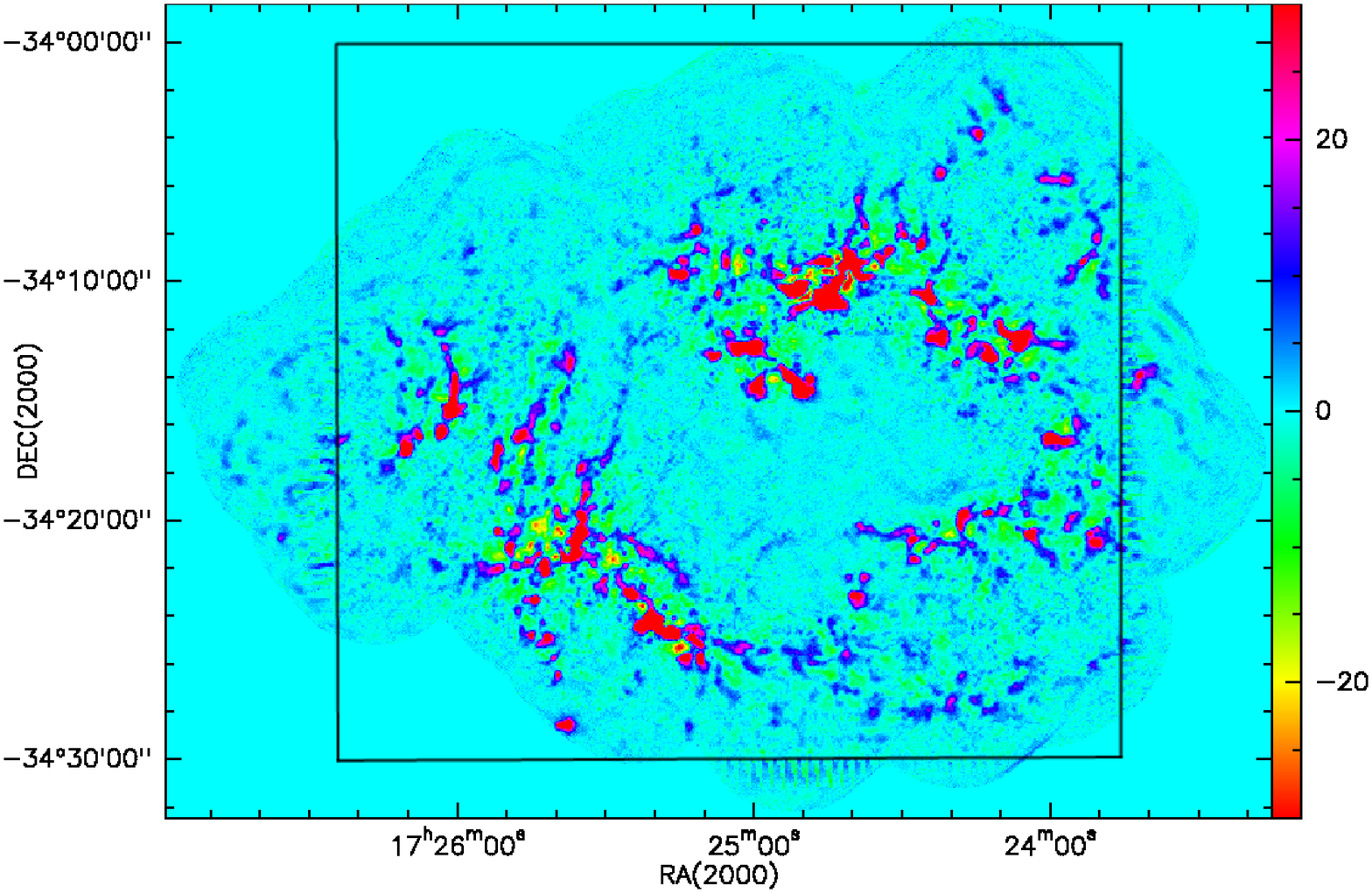}  
\caption{850 $\mu$m map after filtering out emission from extended structures as explained in the text. Units are mJy/beam. The black lines indicate the region mapped in \CO(3-2) 
(see Fig.~\ref{fig:CO32map}).}
\label{fig:filtered_850}
\end{figure}

The $850~\mu$m map has the best combination of sensitivity and angular resolution (see Sect.~\ref{sect:scuba2_obs}), and we used this to identify and extract the cores.
Because we wanted to identify cores in this complex, we needed to isolate the most compact structures, which are more likely to collapse and form a single star or a close multiple system, depending on the level of fragmentation, and are therefore more closely connected with the IMF. 
To facilitate the identification, we first removed the extended $850~\mu$m emission in which the cores are embedded. We therefore performed a multi-scale decomposition of the image, using a median filter \citep[see][]{Belloche+11_aap527_145}.
We used seven levels of decomposition on $S/N$ maps, using only those with scales $\lesssim64\arcsec$ in the image reconstruction, thus filtering out any structure more extended than that. This procedure results in the image shown in Fig.~\ref{fig:filtered_850}.
We note that the spatial resolution at 850 $\mu$m is
$\sim 0.1$ pc, so cores are at best barely resolved. Nevertheless, hereafter we will designate the compact structures found as `cores'.
       
We first constructed a data cube using pyfits, inserting empty two-dimensional planes before and after the actual continuum data, as suggested in \citet{Mookerjea+04_aap426_119}, and modifying the header accordingly; on this data cube we ran Gaussclumps
\citep{1990ApJ...356..513S} as implemented in GREG. 
The algorithm decomposition of the filtered $S/N$ map was then visually inspected to identify and remove spurious sources, confirmed as such by a comparison with the Herschel Hi-GAL maps.
We also excluded cores outside of the region observed in CO with HARP, and at the noisy edges of the continuum mosaic. A minimum size threshold was set at 13\arcsec\ ($\sim$ beam at 850$\mu$m).
The final catalogue contains 1221 cores.
        
\begin{figure}
\includegraphics[width=\columnwidth]{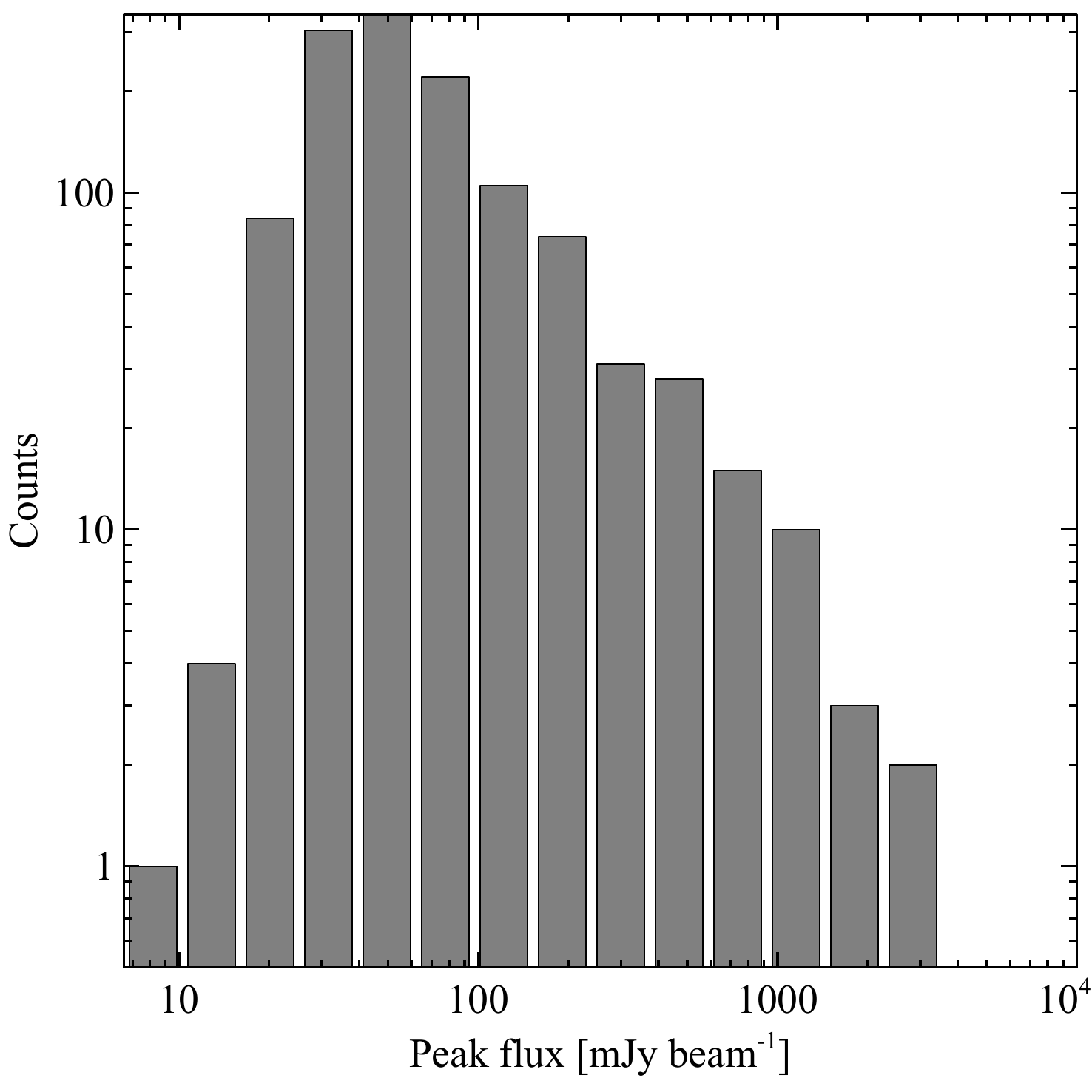}
\caption{Distribution of fluxes for the 1221 cores identified in the $850~\mu$m mosaic. }
\label{fig:flux_distr}
\end{figure}

\begin{figure}
\includegraphics[width=\columnwidth]{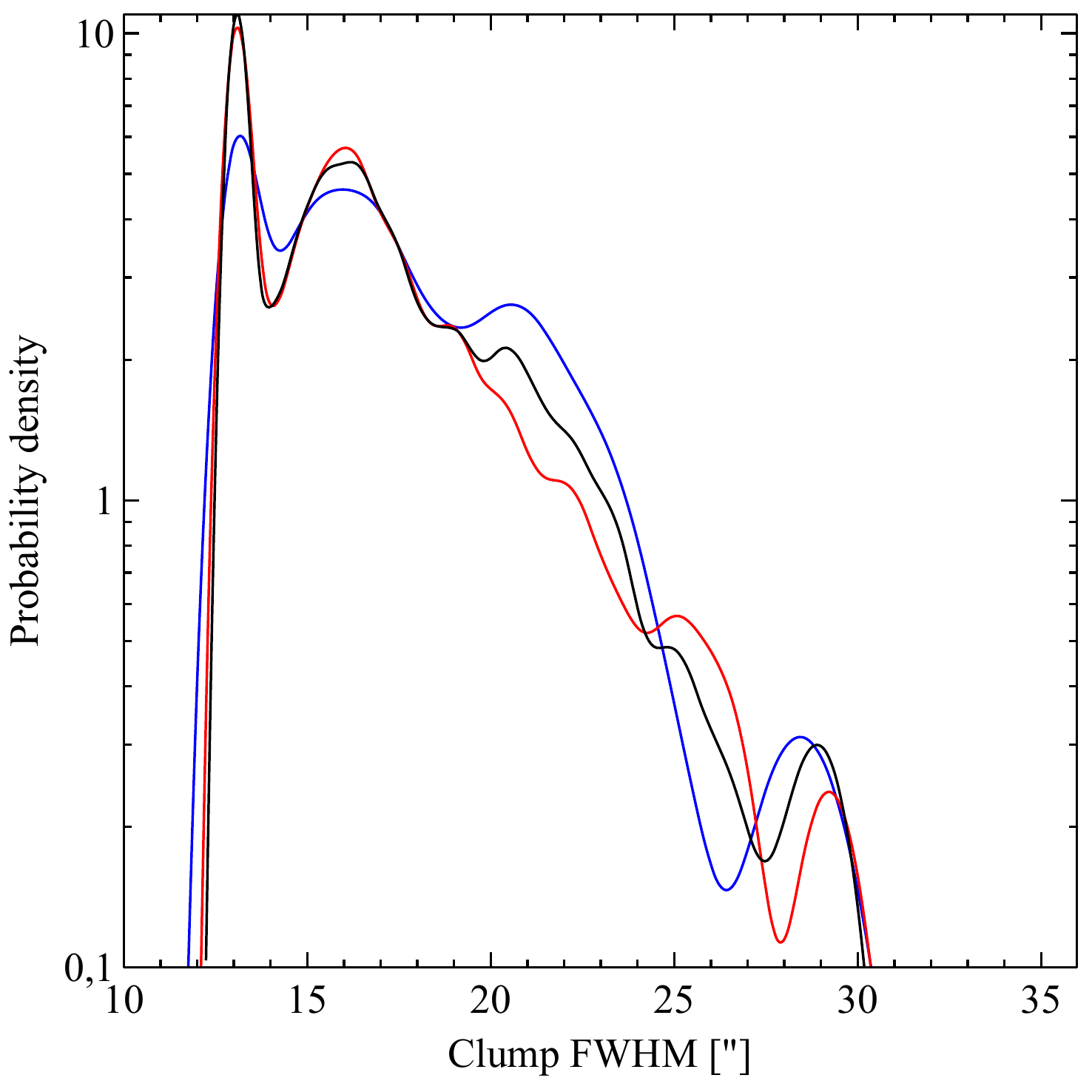}
\caption{Distribution of core sizes for all 1221 cores (uncorrected for beam; black solid line); the red curve indicates the active region and the blue one the more quiescent part. }
\label{fig:size_distr}
\end{figure}

The distributions of fluxes and sizes of the cores in this catalogue are shown in Figure~\ref{fig:flux_distr} and \ref{fig:size_distr}, respectively. In the latter figure, we have also made a distinction between cores in regions with and without intense FUV-flux, as defined in Sect.~\ref{act:qui}.
        
\subsection{'Active' and 'quiescent' regions}
\label{act:qui}
To investigate how the enhanced FUV radiation fields from newly born massive stars affect the molecular environment and the ongoing star formation activity, 
and to avoid effects due to possible dust opacity differences in intense and moderate FUV environments (see Sect.~\ref{dust:opa}), we divided the mapped region into two sub-regions, hereafter the `active' and the `quiescent' regions. The active region includes the parts of the cloud complex more exposed to a high FUV-photon flux, such as G353.2+0.9. On the other hand, the quiescent region includes molecular clouds farther from high-mass stars. Thus, we basically separated the complex into its eastern and western halves (see Fig.~\ref{fig:cores}). The eastern half contains the cluster Pismis 24 (along with the H\textsc{ii} region G353.2+0.9), the cluster AH03J1725--34.4 (along with the H\textsc{ii} region G353.1+0.6), and the cluster B (along with the H\textsc{ii} region G353.2+0.7). This represents the active region. The western half is bound to be less affected by FUV radiation and has been labelled as quiescent region. 
We note that we are only interested in finding out if a global signature of different levels in the FUV is apparent in the core populations in the two areas that may be further investigated in future works. Thus we do not need very accurate selection criteria to define the two regions at this stage. 

\subsection{Spectral energy distributions}\label{sec:SED}
\begin{figure}
\includegraphics[width=\columnwidth]{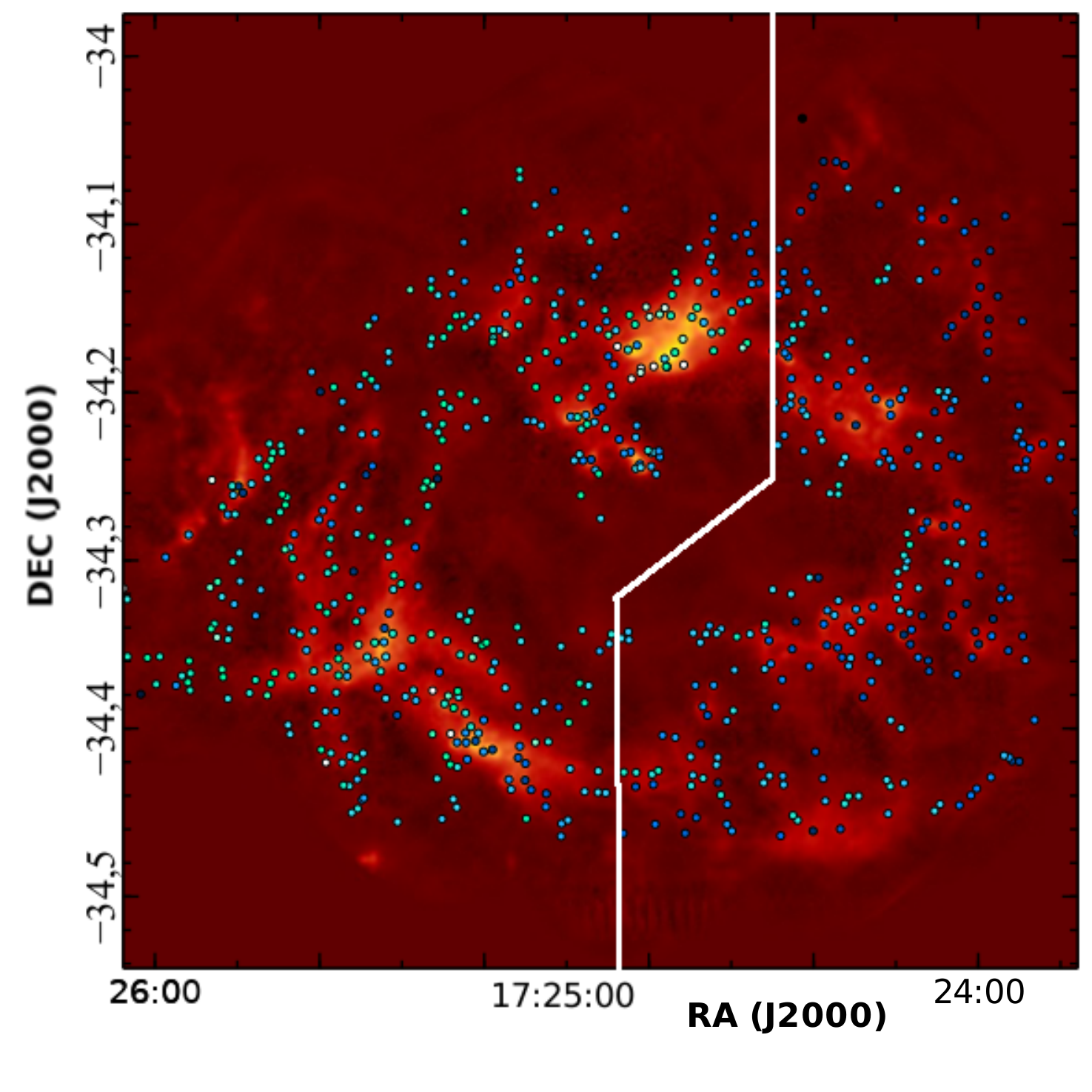}
\caption{All 686 cores with a valid SED fit superimposed on the mosaic at $850~\mu$m. Darker symbols indicate lower temperatures. The white line indicates the dividing line between active (to the east) and quiescent (to the west) regions.}
\label{fig:cores}
\end{figure}

Because we are interested in the CMF, we need to transform the fluxes of the cores into masses. Making use of the Hi-GAL images we constructed the core SEDs by complementing our SCUBA-2 $850~\mu$m and $450~\mu$m flux densities with those from the $160~\mu$m and $70~\mu$m frames taken at the central pixels of the cores identified in the $850~\mu$m map. The Herschel- and the $450~\mu$m maps were first convolved to the 
resolution of the $850~\mu$m map and had their background subtracted in the same way as done for the $850~\mu$m frame.
We fitted a greybody to the SEDs for those cores with at least three data points and a $450~\mu$m flux density larger than the CO-corrected $850~\mu$m one, varying the dust temperature between $5\kel$ and $80\kel$ in steps of $0.5\kel$, and the surface density in the range $10^{-3} - 10^{1.5}\gram\cm^{-2}$, in steps of 0.1 in logarithmic scale. The dust opacity $\kappa_\nu$ was 
assumed to be $0.9\cm^2\usk\gram^{-1}$ at $1.3$ mm \citep{OssenkopfHenning94}, and the index $\beta$ equal to $1.8$, consistent with the dust models used. This yields $\kappa_\nu = 1.85\cm^2\usk\gram^{-1}$ at 870 $\mu$m, the value adopted for the ATLASGAL survey \citep{2009A&A...504..415S}. A more detailed analysis on the adopted dust opacity is reported in Sect.~\ref{dust:opa}    
We were able to derive a dust temperature for 
686 cores; their locations are shown in Fig.~\ref{fig:cores}. The active region contains 411 cores and the quiescent region 275, out of the 686 considered. 
The distribution of their temperatures is displayed in Fig.~\ref{fig:temp:all}. Some examples of SED-fits are shown in the Appendix.
 
An important difference between the cores populating the two regions can be seen in Fig.~\ref{fig:temp:all}. The temperature of cores in the quiescent region peaks at $\sim 25$ K, whereas that of cores in the active region peaks at $\sim 35$ K. Cores in the quiescent region are thus significantly colder than those in the active region.
 Formal errors on the temperature from the greybody fits are nearly always less than 1 K.
This clearly demonstrates the effect of FUV photons on the gas. This result is contrary to that found by \cite{2019ApJ...884...77B}, who do not find evidence of H\textsc{ii} regions affecting the surrounding cores in their study of 38 complexes of ionised and molecular gas. However, their target regions are mostly farther away than 2~kpc and do not contain very massive ionising stars.
        
\begin{figure}
\includegraphics[width=\columnwidth]{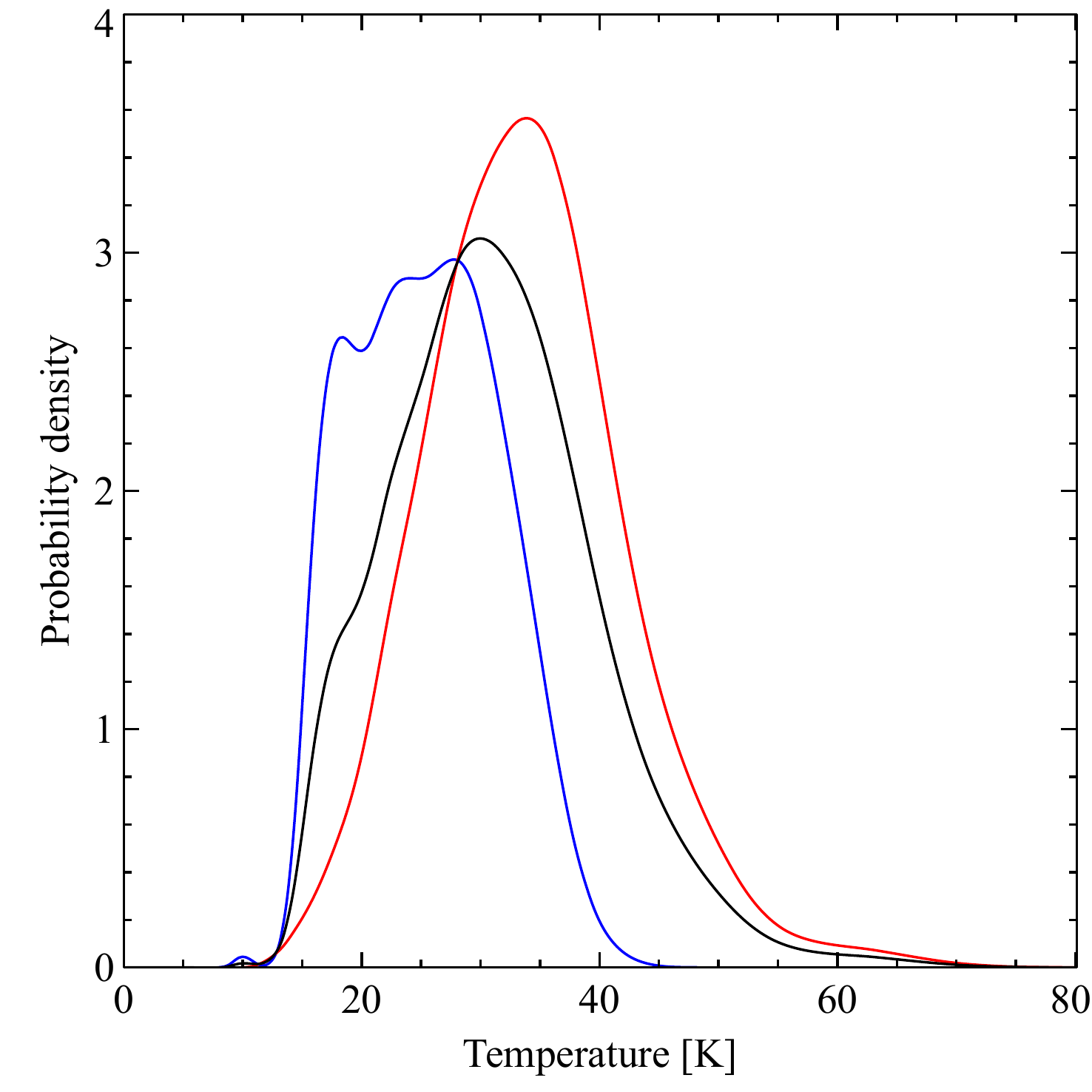}
\caption{Temperature distribution for all 686 cores (solid black line), cores in the active region only (411; solid red line), and cores in the quiescent region only (275; solid blue line). }
\label{fig:temp:all}
\end{figure}
          
\begin{sidewaystable*}
\caption[]{Properties of bona fide cores. Only a few are shown as examples; the full table is at the CDS}
\label{tab:686cores} 
\begin{flushleft}
\begin{tabular}{lccrrrrrcrrrrrr}
\hline
\noalign{\smallskip}
\multicolumn{1}{c}{ID} & \multicolumn{2}{c}{$\alpha$\, \,  (J2000)\, \, $\delta$} & \multicolumn{1}{c}{$F(70)$} & \multicolumn{1}{c}{$F(160)$} & \multicolumn{1}{c}{$F(450)$} & \multicolumn{1}{c}{$F(850)$} & \multicolumn{1}{c}{$S(850)$} & \multicolumn{1}{c}{$\sigma (S(850))$} & \multicolumn{1}{c}{size} & \multicolumn{1}{c}{$M_{\rm gas}$} & \multicolumn{1}{c}{$\sigma (M_{\rm gas})$} & \multicolumn{1}{c}{$T_{\rm d}$} & \multicolumn{1}{c}{$\sigma T_{\rm d}$} & \multicolumn{1}{c}{$\chi^2$} \\ 
 & \multicolumn{1}{c}{degs} & 
 \multicolumn{1}{c}{degs} & 
 \multicolumn{2}{c}{mJy/beam} &
 \multicolumn{2}{c}{mJy/beam} &
\multicolumn{2}{c}{10$^{-24}$\,erg\,s$^{-1}$\, cm$^{-2}$\,Hz$^{-1}$} & \multicolumn{1}{c}{$\prime \prime$} & \multicolumn{2}{c}{M$_{\odot}$} & \multicolumn{2}{c}{K} & \\
\noalign{\smallskip}
\hline
\noalign{\smallskip}
  1 &261.1928& $-$34.1800 & 326569. & 246566. & 21523.7 &  2615.6 &  61.00 &   0.11 & 21.3 & 49.76  & 0.09 &30.5 & 0.3 & 1.3 \\
  2 &261.2131 &$-$34.1720 &  71188. & 140938. & 22403.4 &  2546.8 &  52.00  &  0.10 &19.0 & 60.55  & 0.11 & 23.5 & 0.4 & 2.0 \\
  3 &261.5060& $-$34.2559 & 218380. & 102521. & 48956.8 &  5560.5 &  76.00  &  0.15 &16.2 &175.18  & 0.35 & 15.0 & 2.0& 23.4 \\
  ...& & & & & & & & & & & & & & \\
  ... & & & & & & & & & & & & & & \\
 684& 261.5265& $-$34.3407 &  12696. &6593. &   528.2  &   22.5   & 0.20  &  0.07 & 13.1 &  0.11 &  0.04& 41.0 & 0.0 & 5.7 \\
685 & 261.2487& $-$34.3745  &  3362. & 2009. &$-$10000.0 & 25.6 &   0.22 & 0.07& 13.1 &  0.18 &  0.06 &30.5 & 0.0 & 1.6 \\
686 & 261.3346& $-$34.1587 &   3347. &2516. &   476.3  &   26.2 &   0.42 & 0.12& 17.8 &  0.36 &  0.10 & 29.5 & 0.0 & 3.9 \\
\noalign{\smallskip}
\hline
\end{tabular}
\tablefoot{
  The columns contain the following information: Col.~1: core identification number; Cols.~2 and 3: core position in decimal degrees in RA and DEC; peak flux density \\ at 70, 160, 450 (after convolution to the resolution at 850~$\mu$m and subtraction of the extended emission; $-$10000.0 means no detection) and 850 micron (after subtraction of \\ the extended emission and corrected for $^{12}$CO(3--2) emission) in Cols.~4-7; in Cols.~8 and 9 the total flux density at 850~$\mu$m and its formal (photometric) error (extended \\ emission subtracted and corrected for $^{12}$CO(3--2) emission); in Col.~10: the size (not deconvolved for beam size), which is the geometrical mean of the ellipse axes \\ ($\sqrt{\rm major-axis \times minor-axis}$) at 850~$\mu$m; mass and mass formal error, both derived from the total flux density at 850~$\mu$m, in Cols.~11 and 12; Cols.~13 and 14: dust \\ temperature along with its formal error; $\chi$-square value from the greybody fits in Col.~15.}
\end{flushleft}
\end{sidewaystable*}
        
\subsection{Core masses}\label{sec:mass}
\begin{figure}
\includegraphics[width=\columnwidth]{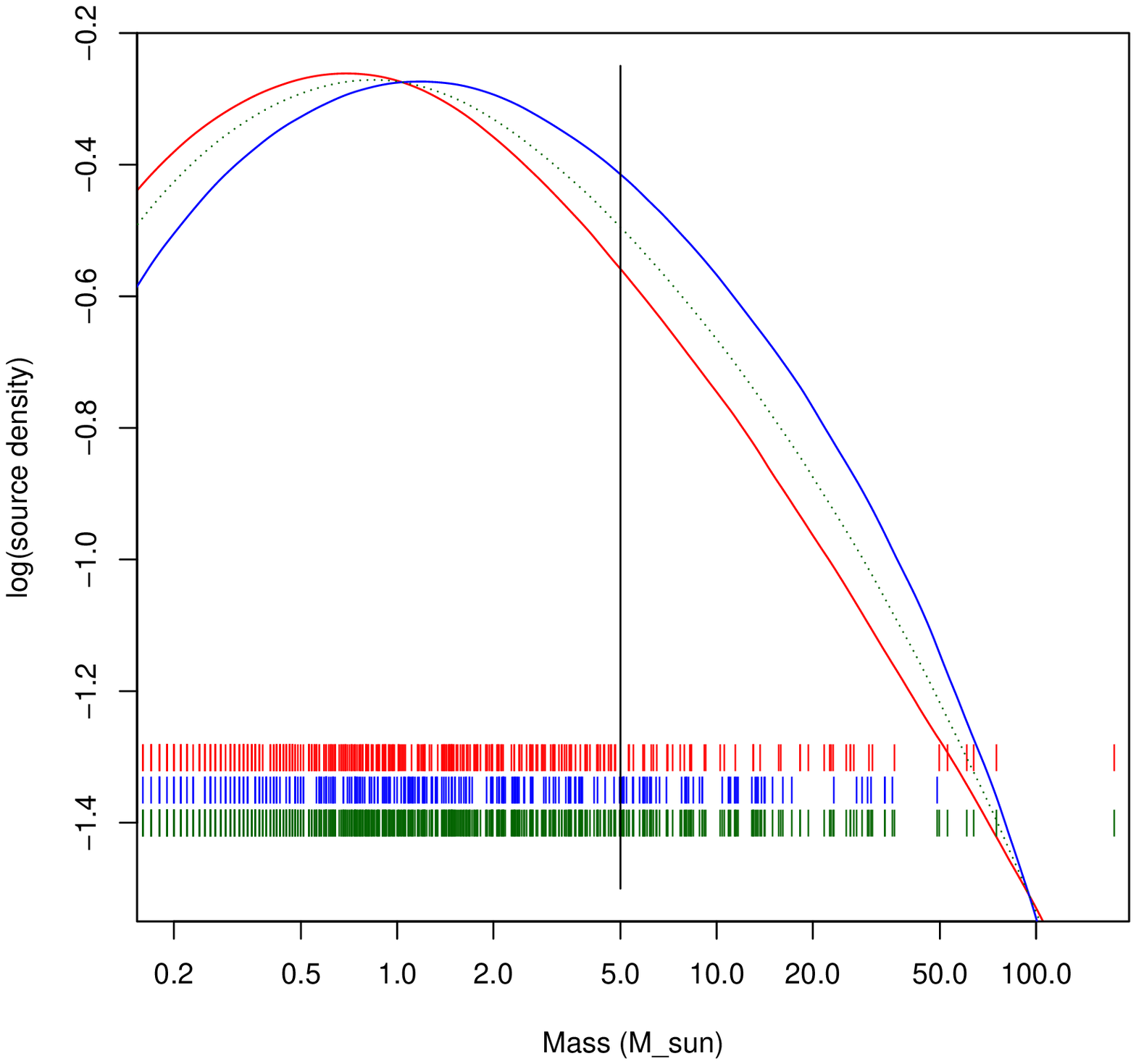}
\caption{Core mass function in NGC6357 (dashed green curve for all cores) obtained as density distribution of sources after smoothing with a Epanechnikov kernel. The red curve indicates the active region, and the blue curve indicates the more quiescent part. The vertical line marks the $5~\msun$  completeness limit (see Sect.~\ref{sec:completeness}). The
small coloured bars show the masses of the single sources for each sample, following the colour coding of the distributions.}
\label{fig:cmf}
\end{figure}

The dust temperatures ($\td$) of the best-fit models are used to calculate the mass of the cores, from the CO-corrected integrated $850\mu$m flux densities from Gaussclumps by means of:
\begin{equation}
\label{equy}
M = \gamma D^2 S_\nu / [\kappa_\nu B_\nu(\td)],
\end{equation}
where D is the distance to the complex, $B_\nu(\td)$ is the Planck function evaluated at temperature $\td$, $S_\nu$ is the total flux density and $\kappa_\nu$ is the dust opacity, both at $850~\mu$m; $\gamma$ is the gas-to-dust ratio (assumed to be 100).
In Table~\ref{tab:686cores} 
we list the properties of all the bona fide cores.
The CMF in the NGC6357 region was constructed through a kernel density estimate, using an Epanechnikov kernel with a bandwidth estimated using the direct plug-in method \citep{2013pss2.book..445F}.
Figure~\ref{fig:cmf} shows the results of this procedure (green dashed line). The slope of the high-mass end of the CMF ($M > 5 \msun$; the completeness limit - see Sect.~\ref{sec:completeness}) is $-2.1\pm0.2$, derived by directly fitting a power-law distribution to the data points using PyMC \citep{Patil+10_jstatsoft35_1}. 

The total gas mass of dense cores is
$\sim 1000$  and $\sim 1500$ M$_{\sun}$ in the quiescent and active regions, respectively. Assuming a molecular mass of
$\sim 1.8 \times 10^{5}$ M$_{\sun}$ for the whole NGC6357 complex (\citealt{2011MNRAS.415.2844C},
re-scaled to 1.7 kpc), that implies a fraction of gas in dense cores of $\sim 1.4$ \%. Such low values are typical in the Galaxy \citep{2016A&A...585A.104C}, reflecting a low SFE (e.g. \citealt{2019ApJ...884...77B}).
        
\subsection{Dust opacity uncertainties}
\label{dust:opa} 
As mentioned above (Sect.~\ref{sec:SED}), we adopted a dust opacity of $0.9\cm^2\usk\gram^{-1}$ at $1.3$ mm
and $\beta = 1.8$, leading to $1.93\cm^2\usk\gram^{-1}$ at $850$ $\mu$m,
based on the theoretical work of
\cite{OssenkopfHenning94}. Their tabulated values are obtained for dense protostellar cores ($\sim 10^{5}-
10^{8}$ cm$^{-3}$) that have remained stable for at least $10^{5}$ years. 
We have also assumed that the dust opacity is constant throughout the sample of cores, so it is essential that
we should check whether this is a good approximation before deriving global properties such as mass distributions
and mass-size relationships.  Although \cite{OssenkopfHenning94} note
that uncertainties on the tabulated values should not add up to more than a factor of 2, one has to take into account 
that most of the cores in NGC6357 have been exposed to an intense FUV field, a scenario that was cautioned but not modelled by \cite{OssenkopfHenning94}.  In addition, from Tables 2 and 3 in 
their paper it is clear that dust opacity depends on the gas density and that its core-to-core variations will be much smaller if ice mantles have grown on the grains. Unfortunately, one can reasonably expect that part of these ice mantles either have been removed or did not grow at all in the presence of a FUV field. We note that systematic errors of even a factor of 2--3 in the dust opacity are acceptable, in that they do not affect global properties of the cores such as the shape of their mass distribution. This is not true for large core-to-core variations in the dust opacity, which in principle may heavily affect the CMF slope.

To test the effects of dust opacity uncertainties in the region, 
one can then assume the worst case scenario of grains with no ice mantles at all.
We derived the dust opacity at 850~$\mu$m as a function of density from Tables 2 and 3 of \cite{OssenkopfHenning94}
by a linear fit in the log-log space. 
For the 452 cores with a deconvolved size of at least half the telescope beam and using the masses derived in Sect.~\ref{sec:mass}, we derive a range of average densities $\sim 5 \times 10^{4} - 10^{5}$~cm$^{-3}$. 
Many cores therefore may have densities less than $10^{5}$ cm$^{-3}$; the
fit allows us to extrapolate down to the lowest values. This may not give the correct dust opacity in those
cases, but it should be good enough for our test, taking into account the limited range spanned by the core densities.

\begin{figure}
\includegraphics[width=\columnwidth]{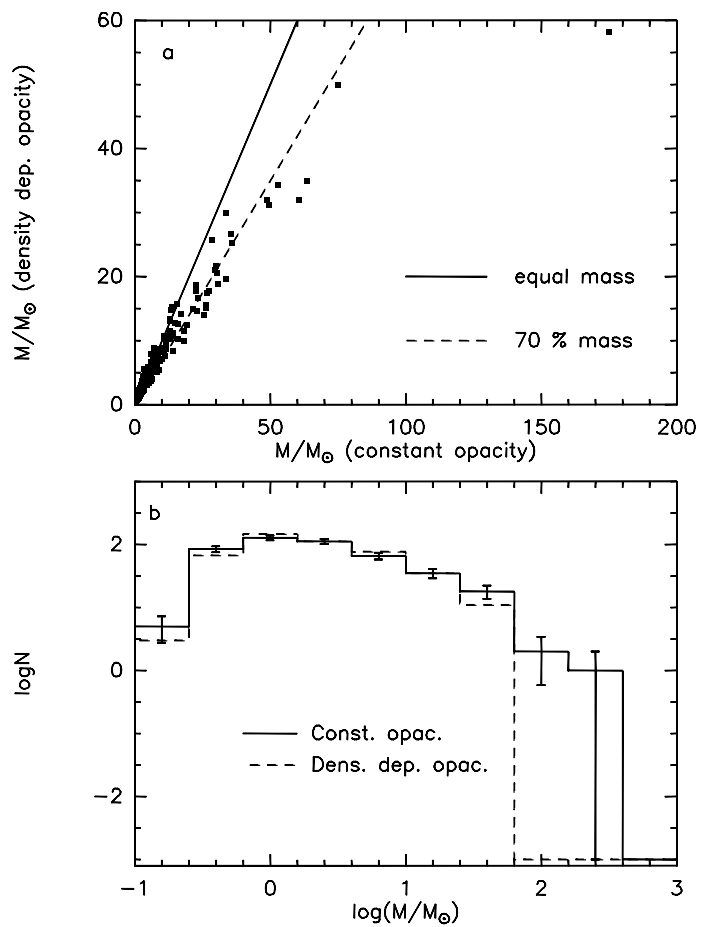}
\caption{Core masses and core mass distribution. {\bf a.} Core masses derived using a density-dependent opacity (grains with no
ice mantles) vs. masses derived using a constant dust opacity of $1.93\cm^2\usk\gram^{-1}$ (at 850~$\mu$m).
The full line marks equal masses, whereas the dashed line marks density-dependent
opacity masses equal to 70 \% of the constant density masses. 
{\bf b.} Core mass distributions for masses obtained with a constant dust opacity
of $1.93\cm^2\usk\gram^{-1}$ (at 850~$\mu$m; full line) and for masses obtained with a density-dependent opacity (dashed line). The error bars have been computed assuming a Poisson    statistics.}
\label{fig:comp:opa}
\end{figure}

Starting from the masses obtained from Eq.~\ref{equy} with the adopted dust opacity
($1.93\cm^2\usk\gram^{-1}$ at 850~$\mu$m), we computed the density of each of the 452 cores and recalculated their mass using the dust opacity for their density, as given by the fitted function.
We then iterated this procedure for each core until convergence was reached. 
The masses obtained by assuming a constant opacity (of $1.93\cm^2\usk\gram^{-1}$) and the ones obtained with a density-dependent dust opacity are compared in Fig.~\ref{fig:comp:opa}a. For most cores, the masses obtained with a density-dependent 
dust opacity are within a $\sim 30$ \% of the masses obtained by adopting a constant opacity. Only three very
massive cores exhibit significant differences (with the density-dependent opacity always yielding lower
masses).

We performed another test by constructing and comparing the mass distributions, 
which are shown in Fig.~\ref{fig:comp:opa}b. These are mostly consistent
with each other within $\sim 1$ $\sigma$ (for a Poisson statistics) and their difference is almost negligible in the 
log-log space, provided the two most massive bins are excluded. According to the mass-size relationships derived in Sect.~\ref{pre:co}, it is easy to see that the most massive cores are also the less dense, so presumably they are more affected by opacity variations.
In fact, a linear fit to the distributions in Fig.~\ref{fig:comp:opa}b (for $M \ge 0.4$ $M_{\sun}$)
yields a slope of $-0.76 \pm 0.06$ for constant opacity and $-0.71 \pm 0.07$ for density-dependent opacity,
in other words consistent with each other within 1 $\sigma$. So the shape of the core mass distribution is only
moderately affected by the choice of the dust opacity. 
A further simple check can be performed to see if the sample masses depend on the size.
We derived mass-size relationships by a linear fit in the log-log space, namely:
\begin{equation}
\label{mass:size:rela}
\log(M/M_{\sun}) = p \times \log(D/{\rm arcsec}).
\end{equation}
We obtained  
$p = 2.35 \pm 0.15$ in the constant opacity case, and $p = 2.49 \pm 0.12$ in the density-dependent opacity case.
Again, the difference is at a $\sim 1$ $\sigma$ level.

We note that these results were obtained by using the average core densities. However, we expect that
significantly higher densities are reached only in a small fraction of a core's total mass.
Furthermore, considering that the density-dependent dust opacity is computed for the worst case of grains with no ice
mantles, the effects of dust opacity uncertainties on the derived masses may be even lower. 
A constant dust opacity is then a good approximation, yielding core-to-core mass uncertainties of $< 30$ \% in most
cases.

\begin{figure}
\includegraphics[width=\columnwidth]{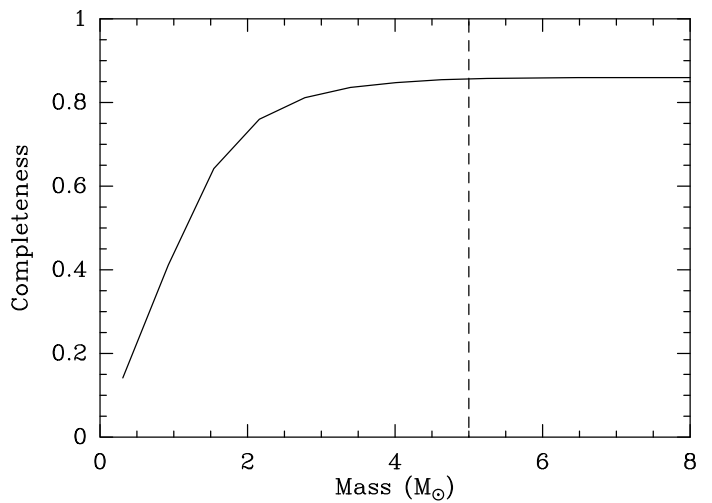}
\caption{Completeness of the cores in our sample, as a function of mass. Starting from the completeness as a function of flux density for point sources, we have used a representative source size of $22\arcsec$ and the observed temperature distribution (Fig.~\ref{fig:temp:all}) to derive the distribution in mass shown here (see text). The dashed line indicates the mass where the completeness reaches its maximum.}
\label{fig:masscompl}
\end{figure}

\subsection{Completeness}\label{sec:completeness}
To evaluate the completeness of our data we followed these steps:
Firstly, we inserted a set of $20$ artificial cores in the $850~\mu$m map with a FWHM of $14\arcsec$ (the SCUBA-2 PSF at $850~\mu$m) 
and a constant peak flux. 
Secondly, the images with the synthetic cores were filtered as done for the original data.
Thirdly, Gaussclumps was run on this processed image and a catalogue of cores was produced, and finally 
this catalogue was cross-matched with the position of the artificial cores. If one or more sources were detected within a beam, we compared their fluxes and sizes, which must be within a factor of 2.5 and 3 
of the input values, respectively, to decide whether the injected source was detected.
This was done for peak flux densities of $5, 10, 20, 40, 60, 80$ and $100\usk\milli\jytab\usk \mathrm{beam}^{-1}$, and the procedure was repeated 100 times for each of these values.
In this way we obtained the completeness as a function of peak flux density.
A completeness of $90\%$ is reached for $60\usk\milli\jytab\usk\mathrm{beam}^{-1}$.
This corresponds to a point-like source with a mass of $M_{\rm compl,point} \sim 0.9$ $M_{\sun}$ for a dust temperature of 20 K and $M_{\rm compl,point} \sim 0.5$ $M_{\sun}$ for a dust temperature of 30 K. 
One way to estimate the mass completeness limits of extended sources would be by assuming a Gaussian spatial distribution of the emission, and then use the distribution over sizes to see how  the completeness of point sources would change;  one can expect that 
\begin{equation}
\label{compl:rela}
M_{\rm compl} \sim M_{\rm compl,point} \times (D^2 + B^2)/B^2, \end{equation}
where $B$ and $D$ are the beam- and deconvolved source sizes, respectively.

We used an alternative method. For each peak flux density value used for the artifical core experiments
(i.e. $5, 10, 20, 40, 60, 80$ and $100\usk\milli\jytab\usk \mathrm{beam}^{-1}$) we computed a cumulative mass distribution from Eq.~\ref{equy} according to the following procedure. Once the peak flux density is fixed, we considered a source size of $\sim22\arcsec$, halfway between unresolved sources ($\sim 13\arcsec$) and the largest (not beam-deconvolved) size values of $\sim 30\arcsec$, and derived the mass distribution obtained from the total flux and the observed distribution of dust temperatures (Fig.~\ref{fig:temp:all}). 
Thus, each peak flux density is associated with a range of masses that reflects the empirical temperature distribution and whose completeness was assumed to be that estimated for this peak flux density from the artificial core experiments described earlier. 
The final mass completeness curve was then computed by averaging the completeness levels derived for each mass value from the adopted  set of peak flux densities, and is shown in Fig.~\ref{fig:masscompl}. 
        
The completeness as a function of mass reaches its maximum around $5\msun$. 
The maximum is not $100\%$ due to confusion effects that prevent the recovery of artificial cores that end up in bright and crowded regions.  We note that our estimates of the mass completeness level assume that the efficiency in retrieving a core mostly depends on its peak flux density, irrespective of its size. This appears reasonable. However, one has to consider that a large fraction of the flux from weak extended cores is likely to be lost in the noise. So the retrieved total flux density from weak extended sources is bound to be affected by larger errors.
                
\section{Discussion}
\label{dis:sec}

\subsection{Comparison with previous works}
\begin{figure}
\includegraphics[width=9cm]{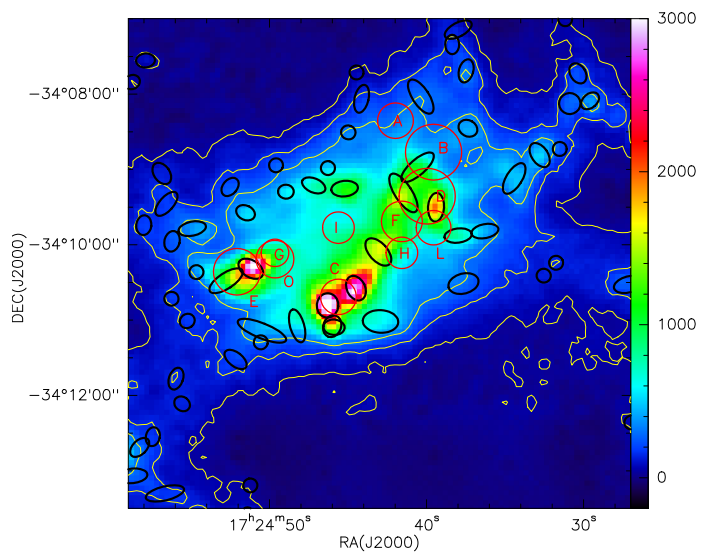}
\caption{Continuum emission at 850 $\mu$m towards G353.2+0.9, overlaid with the locations of the gas cores found by \protect\cite{2012A&A...538A..41G} (red circles
of size equal to that of a clump) and
of the dense cores retrieved in this work (black ellipses). The gas clumps are labelled according to \protect\cite{2012A&A...538A..41G}. Units are mJy/beam. Contour levels are 50, 150, and 300 mJy/beam. } 
\label{gianne12:core}
\end{figure}

The CMF for all cores was derived as discussed in Sect.~\ref{sec:mass} and shown in Fig.~\ref{fig:cmf}. To assess the goodness of both the core retrieval algorithm and the masses obtained, we compare our results to those of previous works. \cite{2012A&A...538A..41G} mapped the area towards G353.2+0.9 in several molecular mm-lines and used data from ATLASGAL \citep{2009A&A...504..415S}, with a beam of $\sim 19\farcs2$ to find the corresponding dust emission in the same region.
They identified 14 clumps from line emission, only three of which with a clear compact counterpart in the continuum emission (their clumps C1, C2, and E).
Figure~\ref{gianne12:core} displays the continuum emission at 850 $\mu$m, overlaid with the positions of the gas clumps found by \cite{2012A&A...538A..41G} and with those of our dust cores. 

Clearly, the increase in spatial resolution has resulted in being able to decompose the continuum emission in a much larger number of compact structures. The goodness of the decomposition operated by Gaussclumps can be confirmed by a visual examination of the figure. Clump C1 corresponds to our core ID 1, clump C2 to our core ID 4, and clump E to our cores ID 2 and ID 20 (see Table~2). We note that \cite{2012A&A...538A..41G} assume a distance of $2.56$ kpc, so their masses have to be scaled to our adopted distance of $1.7$ kpc.

As a first test, we degraded our (line-corrected) 850 $\mu$m map to the ATLASGAL resolution. Then we performed a rough aperture photometry over the areas of the continuum cores found by \cite{2012A&A...538A..41G} (which they label 1 to 5; see their table 3). By assuming the same dust temperature (i.e. 30 K), we retrieve masses that are a factor of $0.5-1.2$ of theirs, confirming that the calibration at 850 $\mu$m is consistent with that of the ATLASGAL data.

As for the gas clumps A to P of \cite{2012A&A...538A..41G}, we found that the masses of our cores are a factor of $0.3-0.5$ of those of the corresponding gas clumps. The gas clumps are colder than our dense cores and using the gas temperature would drive the masses of our cores into accordance with those derived from the gas. 
But it should be noted that using the gas temperature for the dust cores 1 to 5 of \cite{2012A&A...538A..41G} would increase their masses well above those of the corresponding gas clumps.
This indicates that the difference in mass between our cores and the gas clumps is not due to the adopted temperatures, but to the higher spatial resolution and sensitivity of the 850 $\mu$m map, which allow us to retrieve denser and more compact structures embedded in the gas. This is also confirmed by the fact that a few gas clumps are only associated with faint compact continuum sources or faint diffuse emission.

\begin{figure*}
\includegraphics[width=18cm]{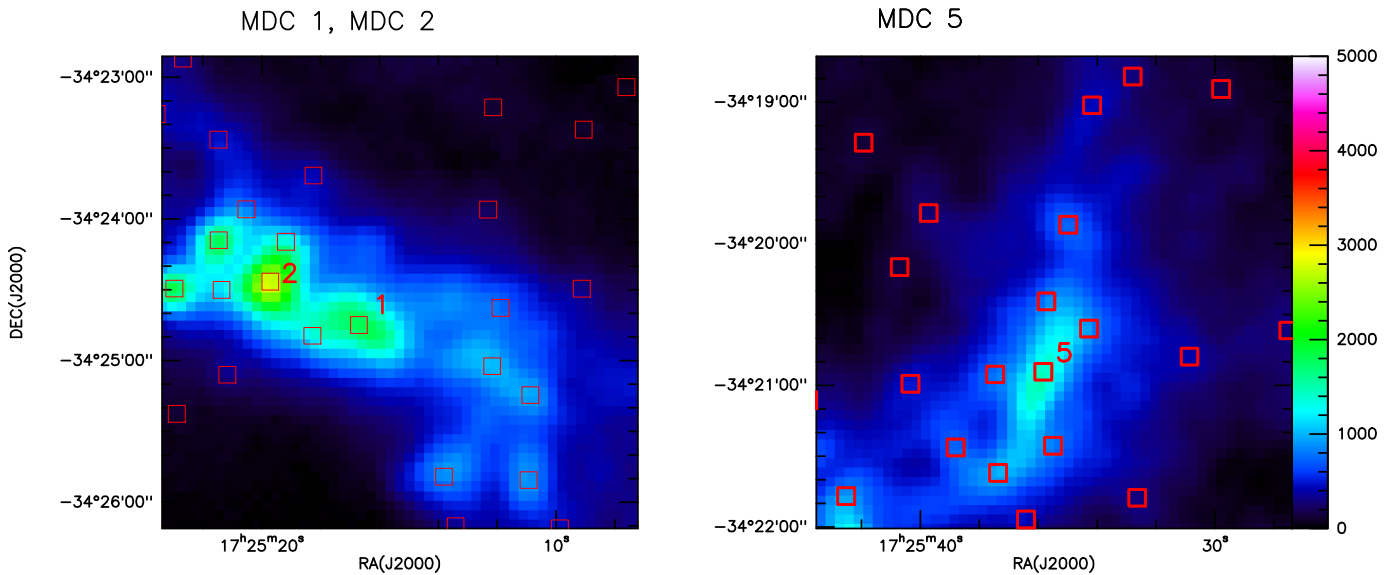}
\caption{Continuum emission at 850 $\mu$m towards the fields of cores MDC 1, 2, and 5 (labelled in figure) of \protect\cite{2019A&A...625A.134R}.
Units are mJy/beam.
The locations of the cores identified in this work are overlaid (small open squares).} 
\label{russe:comp}
\end{figure*}

Our results can also be compared with those of \cite{2019A&A...625A.134R}. These authors used {\it Herschel}/PACS and SPIRE data to select a robust sample of
155 dense cores in NGC6357 with the algorithm {\it getsources} \citep{2012A&A...542A..81M}. The ranges of size and mass that they find
are roughly consistent with ours. However, their mass distribution is peaked around 30--50 $M_{\sun}$, with
a median mass of 22 $M_{\sun}$ and a maximum mass of 386 $M_{\sun}$, while our CMF exhibits a maximum mass of 176 $M_{\sun}$ and a peak at $\sim 1$ $M_{\sun}$. Thus we have achieved a much lower mass completeness limit than \cite{2019A&A...625A.134R}, which is likely to be caused by not only the different algorithms used to decompose the continuum emission but also our preliminary filtering out the more extended emission. We further performed a closer comparison selecting their three most massive sources (their MDC 1, 2, and 5). As shown in Fig.~\ref{russe:comp}, we have retrieved those sources (our cores ID 13, 7, and 14, respectively), along with a number of fainter compact structures around them. The masses of our corresponding cores are a factor of $4-5$ lower than theirs, which cannot be due to the temperatures used or the difference in the adopted dust mass opacity (their prescriptions would yield $\sim 1.24$ cm$^{2}$ g$^{-1}$ at 850 $\mu$m). This clearly indicates that the cores in our sample are systematically less massive than theirs. 
As mentioned above, this difference is likely to arise not only due to the different algorithms used to decompose the emission, but also to our preliminary filtering out lower spatial frequency emission.

\begin{figure*}
\includegraphics[angle=-90,width=12cm]{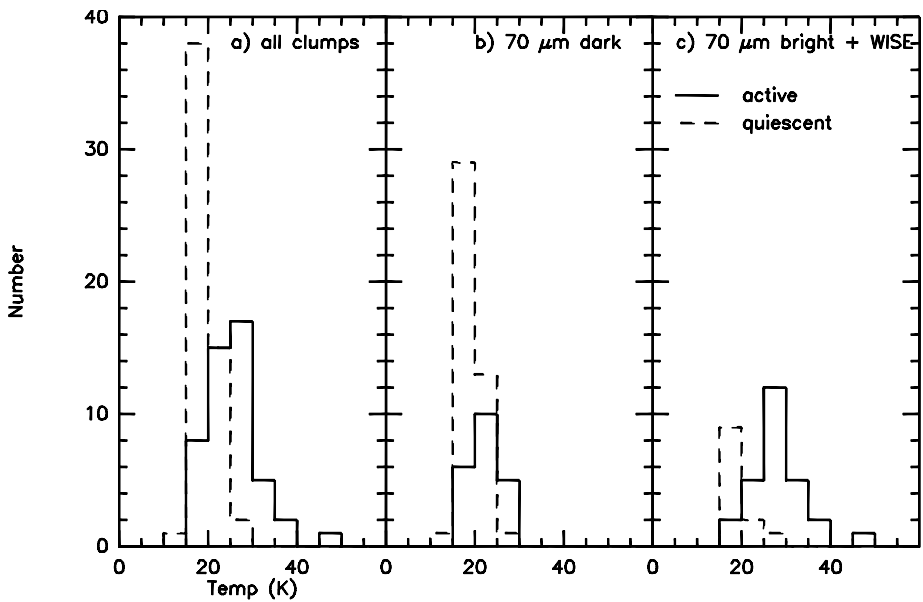}
\caption{Distribution of temperatures for cores in the active (full line) and quiescent (dashed line) regions. Only cores with $M > 5$ $M_{\sun}$ (i.e. above the mass completeness limit) have been included.  
{\bf a.} All cores. {\bf b.} Cores with no emission at 70 $\mu$m and no red WISE sources associated. {\bf c.} Cores associated with 70 $\mu$m emission and/or red WISE sources.}
\label{fig:temperatures}
\end{figure*}

\subsection{Pre-stellar and protostellar cores \label{pre:co}}
Pre-stellar cores are defined as gravitationally bound cores that do not host protostars yet. 
Separating protostellar and pre-stellar cores is critical, in that
the CMF of `protostellar' cores may be affected by mass erosion due to increased 
turbulence and outflows, and the development of internal temperature gradients, and is less suitable to 
compare with stellar IMFs. 
Furthermore, dust opacity is sensitive to the physical environment and
may vary up to a factor of $\sim 2$ in different physical environments, as discussed in Sect.~\ref{dust:opa}. 
It is then essential that the selected sample contains a class of objects
as homogeneous as possible, with a limited range in density. 

 \begin{figure}
\includegraphics[width=9cm]{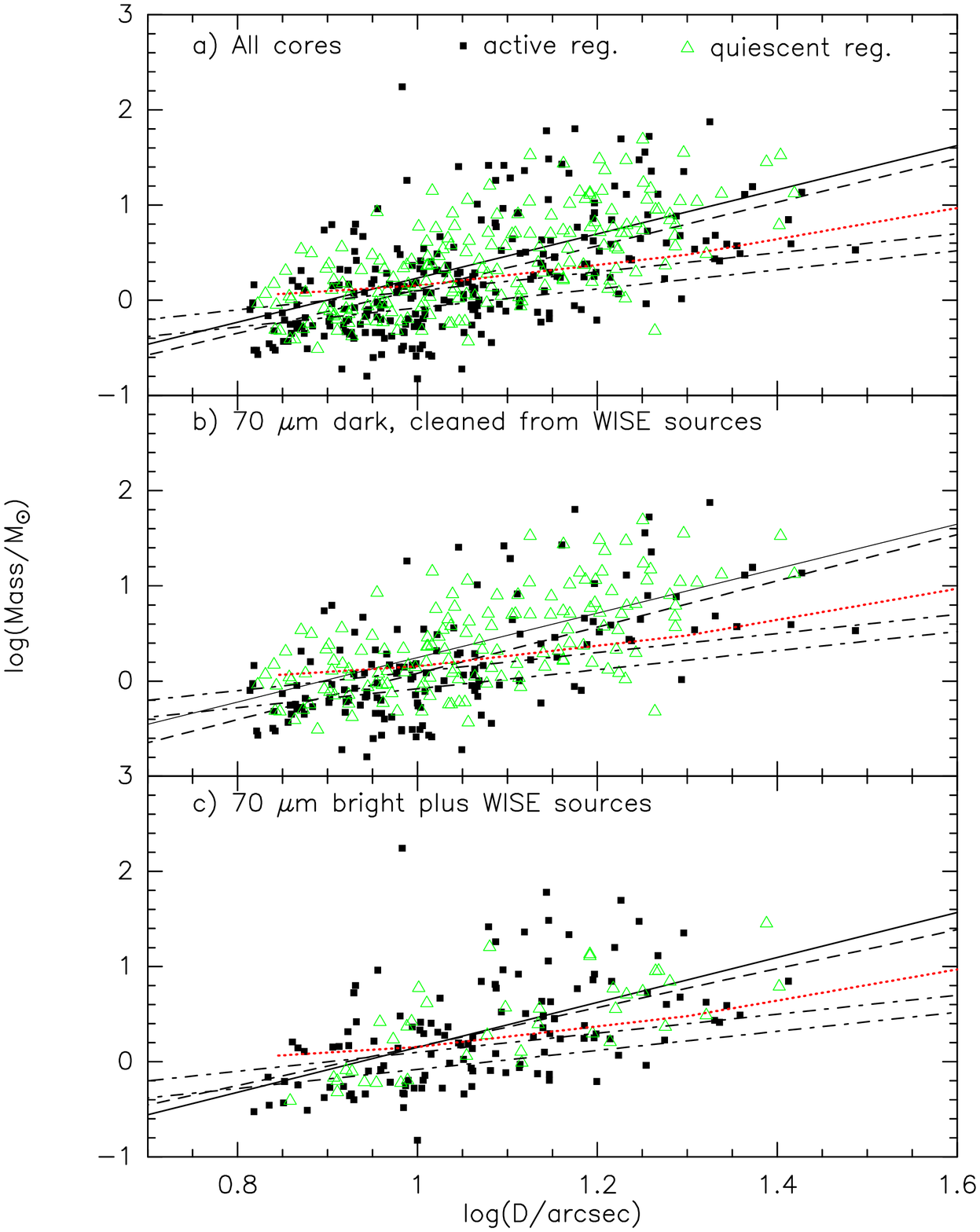}
\caption{Mass vs. deconvolved size for {\bf a.} all cores; {\bf b.} 70 $\mu$m dark cores cleaned from WISE-associated cores; {\bf c.} 70 $\mu$m bright cores plus cores associated with WISE sources.
Cores with a deconvolved size smaller than half the beam size have been discarded.
Small filled squares mark cores in the active region, whereas open green triangles mark cores in the quiescent region. The critical half-mass loci of Bonnor-Ebert spheres are drawn for $T = 20$ K and $T = 30$ K (black dash-dotted lines). The dotted red line marks the completeness limit for $T = 20$ K dust (from Eq.~\ref{compl:rela}). Linear (log-log) fits are indicated by a solid (cores in the quiescent region) and a dashed (cores in the active region) line.}\label{fig:stability}
\end{figure}
        
Identifying protostellar cores 
requires probes of protostellar activity, such as compact emission at 70~$\mu$m.  Thus, we cross-checked our 850~$\mu$m data and Herschel Hi-GAL 70 $\mu$m maps for spatial coincidences. All sub-mm cores associated with 70~$\mu$m emission have 
then been labelled as protostellar,  whereas all remaining cores have been labelled as `starless'.  
It has been shown that 70~$\mu$m emission is a sensitive probe of protostellar activity (see e.g. \citealt{2012ApJ...744..130K}). 
The distribution of the peak flux densities of the 70~$\mu$m emission associated with the sub-millimetre cores, has a maximum at $\sim 4$~mJy/beam. This represents a rough completeness limit. Assuming that the infrared emission comes from a Gaussian spatial distribution $\sim 20\arcsec$ in size and a FWHM $\sim 9\arcsec$ for
the Herschel beam at 70 $\mu$m, this yields a flux of $\sim 24$~mJy. Then by using Eq. 2 of \cite{2008ApJS..179..249D} one can estimate a detection limit for the luminosity of the central protostar $L_{\rm int} \sim 0.3$ $L_{\sun}$. 
One problem with this approach is that in the active region, dust is heated from the outside as well, through FUV radiation, thus increasing the core emission at 70~$\mu$m possibly mimicking the effects of a central protostar.
In fact, we found 129 `70~$\mu$m-bright' 
sources out of 411 cores in the active regions, and only ten 70~$\mu$m-bright sources out of 275 cores in the quiescent region. 
All 70~$\mu$m-bright cores in the active regions are associated with one of the two H\textsc{ii} regions G353.2+0.9 and G353.1+0.6. This indicates a possible bias due to far-infrared 
emission from warm dust 
in the outer parts of cores located inside or near to the H\textsc{ii} regions.

We then applied a further selection filter to the 70~$\mu$m-dark cores, based on the spatial coincidence of cores and red WISE sources. After retrieving all point sources from the WISE Source 
Catalog\footnote{http://wise2.ipac.caltech.edu/docs/release/allsky/\#src\_cat} projected towards NGC6357, we selected those with colours of Class I sources according to \cite{2012ApJ...744..130K}. All contaminants (extragalactic sources, 
PAH emission, shock emission) were discarded following the criteria of \cite{2012ApJ...744..130K}. 
We also included sources detected in only two bands with ${\rm d}\log(\lambda F_{\lambda})/{\rm d}\log\lambda 
> -0.3$ and sources detected at only 22~$\mu$m. In addition, we required photometric errors $< 0.3$ mag in 
all the relevant bands. Then we cross-correlated this cleaned list of infrared sources and the one of sub-mm cores. 
All 70~$\mu$m-dark cores whose ellipse overlaps the positional uncertainty ellipse of a WISE source were re-classified as protostellar; this concerned 31 cores in the active region and 41 in the quiescent region. In the end, we found 251 starless and 160 protostellar cores in the active region and 224 starless and 51 protostellar cores in the quiescent region. 
In principle we could also use the WISE sources to verify the nature of the 70~$\mu$m-bright cores in the areas of the active region closest to the H\textsc{ii} regions and in the PDR. 
Unfortunately this is impossible, due to the saturation of the WISE bands; 
therefore in those areas all 70~$\mu$m-bright cores remain tentatively classified as protostellar.

In Fig.~\ref{fig:temperatures} we show the temperature distributions for the various (sub)samples. To avoid effects due to incompleteness, only cores with $M > 5$ $M_{\sun}$ have been selected. Fig.~\ref{fig:temperatures}a shows the distribution for all cores, and emphasizes what we saw in Fig.~\ref{fig:temp:all} for cores of all masses. The temperature distributions for 70~$\mu$m-dark and 70~$\mu$m-bright plus WISE-associated cores in the active and quiescent regions are shown in Figs.~\ref{fig:temperatures}b, c. The temperatures in the quiescent region for both classes are similar, peaking between 15--20~K with an average $T \sim 19$~K. On the other hand, in the active region the distribution of 70~$\mu$m-bright cores is clearly shifted to higher temperatures with respect to that of 70~$\mu$m-dark cores (from an average $T \sim 22$~K to an average $T \sim 28$~K).

The sensitivity limit for protostellar cores associated with WISE sources can be estimated as follows. We will assume here that
the central protostellar source has an infrared flux increasing with wavelength, with a spectral index ${\rm d}\log(\lambda F_{\lambda})/{\rm d}\log\lambda \sim 2$. This is typical of embedded Class I sources. As the
flux is higher in the WISE band at 22 $\mu$m, we need to derive the WISE sensitivity to point sources in this band.
The $5\sigma$ sensitivity in the 22~$\mu$m band quoted in the explanatory 
supplement\footnote{http://wise2.ipac.caltech.edu/docs/release/allsky/expsup/} is $5.4$~mJy (i.e. the magnitude at 22~$\mu$m ($[22]) \sim 8$). As the flux at 22 $\mu$m and the spectral index are known, we can compute the flux that such a source would exhibit in the other WISE bands.
Following \cite{2012AJ....144...31K}, who relate infared spectra to central bolometric luminosities, $F_{22} \sim 5.4$~mJy and spectral index 2 correspond to 
a protostar of $\sim 3.5$~$L_{\sun}$. However, WISE is much more sensitive in the other bands, thus central sources with 
the same spectral index can be undetected at 22~$\mu$m but detected in the lower wavelength bands, increasing the actual sensitivity. On the other hand, due to saturation in the WISE bands
there is in fact no way to discriminate between starless and protostellar cores in the areas towards the two 
H\textsc{ii} regions, that is to say the most critical ones to understand the effects of FUV feedback. Even using the MSX Point Source Catalogue, which has a worse spatial resolution and sensitivity but saturation at higher infrared fluxes, yields only 20 matches near to the H\textsc{ii} regions, still indicating sensitivity problems. All cores associated with MSX point sources have temperatures in the range 26--63~K, with only two cores above 5 $M_\sun$ ($T = 26$ and $34.5$~K). 
Moreover, the histogram plotting the number of protostellar sources versus $[22]$ displays peaks at $[22] \sim 1.5$~mag and $\sim 2.5$~mag in the active and quiescent regions, respectively, indicating that the actual completeness limit is much worse than inferred from the nominal sensitivity limit.  

A rough analysis of the CMF shows that little is changed in the statistical properties of the population 
of starless cores when excluding those associated with WISE sources. Binning the CMF in intervals of $0.4$ in logarithmic masses and fitting the high-mass end ($\log[M/M_{\sun}] > 0.8$) with a
$N \sim M^{\alpha}$ function, one obtains $\alpha = -1.79 \pm 0.09$ for 70~$\mu$m-dark cores 
and $\alpha = -1.71 \pm 0.1$ for 70~$\mu$m-dark cores cleaned of WISE-associated sources in the quiescent regions.
In the active regions, one obtains $\alpha = -1.72 \pm 0.09$ for 70~$\mu$m-dark cores and $\alpha = -1.62 
\pm 0.12$ for 70~$\mu$m-dark cores cleaned of WISE-associated sources. The difference is not statistically significant, probably due to the fact that only a few sources are moved between samples if their mass is above the completeness limit. 

The discussion above highlights the difficulties inherent in obtaining a census of real starless cores. 
Although 70~$\mu$m emission is in principle a very sensitive probe of protostellar emission, nevertheless we found many cores associated with red WISE sources and no emission at 70~$\mu$m. 
In addition, we have no way of checking the nature of 70~$\mu$m-bright cores around the two H\textsc{ii} regions, which are 
the sources most affected by the FUV feedback.
On the other hand, we have shown that the association with red WISE sources is not a sensitive tool in the whole region,  because of saturation. As a result, there is currently no way of unambiguously separating starless and protostellar cores in large areas towards NGC6357.

For the sake of simplicity, we will refer to the sample of 70~$\mu$m-dark cores cleaned of those WISE-associated as the 'starless' sample, despite its uncertain degree of contamination from protostellar cores. It is composed of 224 sources in the quiescent region and 251 sources in the active region. On the other hand, there are only 51 candidate protostellar cores in the quiescent part and 160 possible protostellar cores in the active region (31 cores that are associated with red WISE sources outside the H\textsc{ii} regions 
plus 129 70~$\mu$m-bright cores).

Not all starless cores are also pre-stellar. Some are just transient, short-lived structures 
that will dissipate rather than form stars. To check how much our sample of starless cores would also be
representative of pre-stellar cores we followed the analysis of \cite{2014prpl.conf...27A}.
In Fig.~\ref{fig:stability}, core masses and sizes can be compared to the half-mass loci of Bonnor-Ebert critical mass for $T = 30$~K and $T = 20$~K. It is clear that cores with masses larger than the completeness limits derived from the sub-mm emission, cannot be thermally supported against gravity. Even though turbulence and magnetic fields can oppose gravitational collapse, it is unlikely that they will prevent it. So, most of our starless cores, if really starless, would also be pre-stellar in nature, provided that only masses above the completeness limit are taken into account.

Mass and size of all cores in the two regions are plotted in Fig.~\ref{fig:stability}a. We note that only cores with deconvolved sizes greater than half the beam size have been used. A linear fit (in log--log space) to the data according to Eq~\ref{mass:size:rela} yields $p = 2.3 \pm 0.2$
both for the active and the quiescent region. This is consistent with $\log(M/M_{\sun}) \sim 2 \times \log (D/{\rm arcsec})$, typical of core populations with constant column density. 
Similarly, a linear fit to the mass versus deconvolved size of starless cores (shown in Fig.~\ref{fig:stability}b) 
yields $p = 2.4 \pm 0.3$ in the active region and $p = 2.3 \pm 0.2$ in the quiescent region. 
The mass-size relations obtained from the linear fit to protostellar cores (shown in Fig.~\ref{fig:stability}c) are quite close to those: $p = 2.0 \pm 0.3$ in the active region and $p = 2.4 \pm 0.4$ in the quiescent region. The differences in the power-law indices are clearly not significant. In addition, one has to be aware that the fit results may be biased due to size-dependent incompleteness, (see Sect.~\ref{sec:completeness}). Nevertheless, the mass-size relationships derived in various star-forming regions or from various samples of cores or clumps displays power-law indices in the range $1.7-2.2$ and somewhat depending on the  algorithm used, as discussed by \cite{2019A&A...628A.110M}, hence similar to the value found in NGC6357.

\subsection{Nature of the cores in the PDRs}
\label{gorti:pdr}
Although we were not able to reliably separate protostellar and starless cores in the molecular gas adjacent to the H\textsc{ii} regions, nevertheless their nature can be assessed if we assume that those cores are embedded in PDRs. As we said, all 70~$\mu$m-bright cores in the active region lie near either G353.2+0.9 or G353.1+0.6. Thus, we applied to this sample the models developed by \cite{2002ApJ...573..215G}. 
These authors investigated the evolution of FUV-illuminated clumps (we note that they designate cores as `clumps') in PDRs by examining both the case of turbulent clumps in a vacuum and that of pre-existing clumps in a confining interclump medium (ICM). 
They found that clump evolution in a PDR is driven by $\eta_{\rm c0}$ (the ratio of the initial clump column density to that of the FUV-heated clump surface, $\sim 10^{21}$ cm$^{-2}$), $\nu$ (the ratio of sound
speed at the surface to that in the innermost clump regions), and the turn-on time of the heating flux
(compared to the initial sound crossing time in the clump).

We computed the deconvolved size for each of the 160 70~$\mu$m-bright cores in the active region and discarded all those smaller than
half the beam size. We then determined volume and column density for each of the remaining 102 sources.
These were further divided into two subsamples, based on their location either towards G353.2+0.9 or
G353.1+0.6.
The FUV field (G$_{0}$) impinging on each core, in Habing units of $1.6 \times 10^{-3}$~erg cm$^{-2}$
s$^{-1}$ was derived from Eq.~14 of \cite{2012A&A...538A..41G} by scaling to a distance of $1.7$~kpc
and an inclination of the ionising sources to the core direction of $45\degr$. We adopted a total FUV stellar luminosity of $1.2 \times 10^{6}$~$L_{\sun}$ for G353.2+0.9 \citep{2012A&A...538A..41G} and
scaled the FUV field derived by \cite{Massi+97} to the estimated star-core distance for G353.1+0.6.
We did not account for extinction of the FUV field and shadowing for the outermost cores, we just assumed that the clumpiness of the gas
allows the radiation to penetrate deeply in the molecular clouds bordering the
H\textsc{ii} regions. Furthermore, the relatively high adopted inclination of $45\degr$ may partly compensate for extinction. 

\begin{figure}
\centering
\includegraphics[width=\columnwidth]{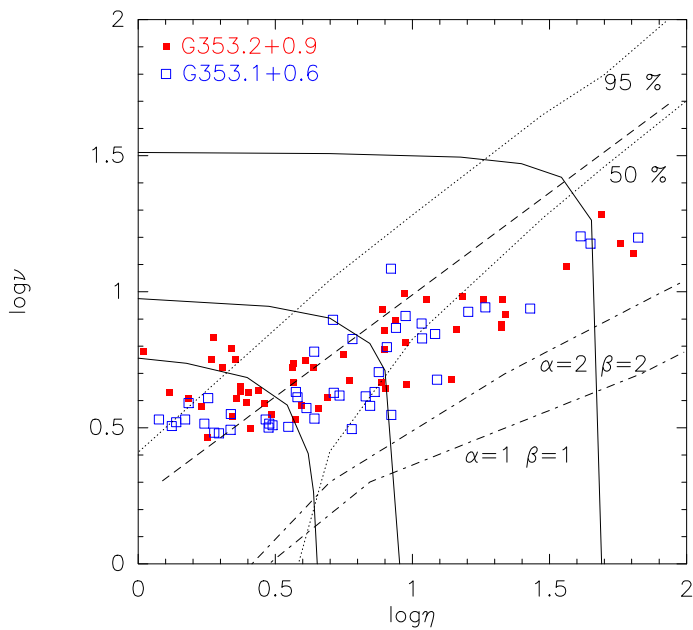}
\caption{$\eta$-$\nu$ (see text) plot for the cores associated with the PDRs around
the H\textsc{ii} regions G353.2+0.9 (filled red squares) and G353.1+0.6 (open blue squares).
The dash dotted lines mark the critical densities for turbulent cores
for different values of the non-thermal to thermal pressure ratio ($\alpha$)
and the magnetic to thermal pressure ratio ($\beta$), and a frozen magnetic field.
The solid lines mark the evolution of the column density ratio $\eta$ against the
parameter $\nu$ (which increases with increasing G$_{0}$) for pre-existing
cores embedded in an interclump medium. The dotted lines indicate the fraction
of mass lost by such cores due to photoevaporation when exposed to a steadily increasing G$_{0}$. The
dashed line marks the critical density for such cores, i. e. the column density
of a core that would be completely photoevaporated by a FUV field $\nu$.}
\label{eta:nu}
\end{figure}

From G$_{0}$ and the average density we derived the core surface temperature 
using the PDR model of \cite{1999ApJ...527..795K}. In turn, we derived $\nu$ for
each core from the surface and far-IR temperatures. It should be noted that $\nu$ measures the
intensity of the FUV field. The dimensionless column density parameter $\eta$,
namely the ratio of the current core column density to that of the FUV-heated core surface ($\sim 10^{21}$~cm$^{-2}$),
was also computed from the mass and the deconvolved size of each core. The results are
shown in Fig.~\ref{eta:nu}, adapted from  diagnostic diagrams from  \cite{2002ApJ...573..215G} 
(see e. g. their Figs.~2 and 7).

\cite{2002ApJ...573..215G} consider two classes of cores populating a cloud
PDR. One class is represented by turbulent cores in a vacuum that form
quickly in the PDR in a time less than that needed by the IF to cross the cloud PDR. It should be noted that these authors do not account for gravity in
their analytical treatment. Assuming
a PDR thickness of $\sim 1$ pc and an IF speed of $\sim 1$ km s$^{-1}$, the lifetime of these cores would be $< 1$ Myr. According to \cite{2002ApJ...573..215G},
these cores (if massive enough) are impulsively heated by the FUV field and would rapidly attain a constant column density irrespective of their initial column density. This critical
column density only depends on $\nu$, the non-thermal to thermal pressure ratio ($\alpha$), and the magnetic to thermal pressure ratio ($\beta$); dot-dashed lines in Fig.~\ref{eta:nu} mark the loci of critical density for $\alpha = \beta =1$ and $\alpha = \beta =2$, see Eqs.~6-7 of \cite{2002ApJ...573..215G}.  

The second class these authors consider are pre-existing cores. 
Pre-existing cores much smaller than the thickness of the cloud PDR and embedded in an ICM would enter gradually the PDR and would experience a steadily
increasing FUV field (and $\nu$). The solid lines in Fig.~\ref{eta:nu} mark the evolution of the column density parameter $\eta$ against the parameter $\nu$, which will steadily increase as
the core moves deeper into the PDR (Eqs. 24--25 of \citealt{2002ApJ...573..215G}). 
The core experiences photoevaporation and the fraction of mass lost during the evolution is indicated by dotted lines in the figure. The dashed line marks the gas column density that would be completely photoevaporated given a FUV field $\nu$ (Eq.~26 of \citealt{2002ApJ...573..215G}). 

The cores in the PDRs around G353.2+0.9 or G353.1+0.6 show the same distribution and are clearly more consistent with pre-existing cores embedded in an ICM than turbulent cores. Although the value
G$_{0}$ has not been corrected for extinction, we note that the highest $\nu$ values always derive from the highest values of G$_{0}$. Therefore, these cores represent the ones nearest to the ionising sources, which are the least affected by extinction. In the density range
spanned by the rest of the sample, the core surface temperature is not much affected by
G$_{0}$ and even a decrease of an order of magnitude would result in a decrease in temperature by a factor of 2--3. Therefore, $\log(\nu)$ would decrease at most by $\sim 0.24$. 

The cores appear to cluster around the critical density for pre-existing gas structures or lie below it in Fig.~\ref{eta:nu}. This suggests that when the increasing $\nu$ (i.e. the IF moves closer and closer to the core) causes the eroded gas column density to decrease below the critical density, a core will be quickly photoevaporated. However, this is inconsistent with the cores with the highest column density displaying the highest values of $\nu$. In fact, this would imply that the cores with the highest column density are located in regions where G$_{0}$ is the highest, that is to say close to the IFs. But there is no reason why equally high column densities should not be found closer to the bulk molecular gas (exhibiting low values of $\nu$).
Then, either turbulence and/or gas compression becomes important close to the IFs and new cores are formed or, more likely, lack of spatial resolution artificially merges smaller cores increasing the measured column density (we note that the IF is seen edge-on towards both H\textsc{ii} regions). 
  
This analysis suggests that the PDRs preceding the IFs penetrate into the molecular clouds through a population of pre-existing molecular cores embedded in a lower density ICM.
Some of these cores are then likely to experience an RDI with a subsequent gravitational collapse (see \citealt{2011ApJ...736..142B}, and references therein).  
This is consistent with the elephant-trunk structures, some associated with IR sources, observed in the region (\citealt{Massi+15_aap573_95}). \cite{2019MNRAS.487.3377D} numerically
modelled the photoevaporation of Jeans-unstable molecular cores, hence including gravity, finding that Jeans-unstable cores will remain Jeans-unstable after the RDI, although losing part of their mass. This indicates that the most massive cores in the cloud PDRs towards NGC6357 should collapse and host forming stars, rather than be just photoevaporated. 
In other words, accelerated or even triggered star formation is expected around the H\textsc{ii} regions.

\subsection{The CMF and comparison with the IMF}

We derived the CMF for the samples of cores in the active and quiescent region and estimated the slope of the high-mass end of the CMF, $\alpha = - {\rm d}\log({\rm d}N/{\rm d}M)/{\rm d}\log(M)$,
using only cores with $M \ge 5$ $M_{\sun}$. 
To minimise statistical biases, which affect simple linear fits in the log-log space (see \citealt{2009MNRAS.395..931M}),
we used the python task PyMC and the maximum likelihood estimator (ML) discussed by \cite{2009MNRAS.395..931M}. The results are listed in Table~\ref{table:alpha}.   

%
%
\begin{table*}
\caption{Power index, $\alpha = -{\rm d}\log({\rm d}N/{\rm d}M)/{\rm d}\log(M)$, 
of the CMFs. Only cores with $M \ge 5$ $M_{\sun}$ have been used.}
\label{table:alpha}
\centering
\begin{tabular}{c c c c c c c}     
\hline\hline
 & \multicolumn{3}{c}{Quiescent area} & \multicolumn{3}{c}{Active area}\\
Sample & PyMC fit & ML & Number of  & PyMC fit & ML & Number of \\
       &          &    & datapoints  &     &    & datapoints \\
\hline 
All cores & $2.4 \pm 0.2$ & $2.1 \pm 0.2$ & 57 & $1.9 \pm 0.1$ & $1.7 \pm 0.2$ & 48 \\
Protostellar cores\tablefootmark{a} & $2.8 \pm 0.5$ & $2.4 \pm 0.6$ & 12 
    & $2.0 \pm 0.2$ & $1.9 \pm 0.2$ & 27\\  
Pre-stellar cores\tablefootmark{a} & $2.3 \pm 0.2$ & $2.0 \pm 0.2$ & 45
      & $1.8 \pm 0.2$ & $1.22 \pm 0.05$ & 21 \\
\hline 
\end{tabular}
\tablefoot{
\tablefoottext{a}{See the text on the problems to obtain actual protostellar and pre-stellar
  samples}
}
\end{table*}

Three trends can be pointed out from Table~\ref{table:alpha}: i) In each region (quiescent
and active), all samples and estimates are consistent with one another within $1 \sigma$,
except for the values for pre-stellar cores in the active area yielded by the ML;
ii) all estimates in the quiescent area are consistent with a Salpeter IMF
($\alpha = 2.35$) within $1\sigma$; iii) all estimated CMFs in the active area are
flatter than in the quiescent area at least at a $2 \sigma$ significance level. Any effect of the FUV field on the dust opacity, discussed in Sect.~\ref{dust:opa}, should not significantly affect $\alpha$ in
the pre-stellar samples (except  possible global shifts in mass, different for each region), which do not include cores embedded in the PDRs surrounding
G353.2+0.9 and G353.1+0.6. We also note that the CMFs in the quiescent area displays slopes $\alpha$ that are consistent with those of the IMFs derived by \cite{Massi+15_aap573_95} for Pismis~24, adopting the smaller extinction range. When the cluster members are selected in the extinction range $A_{\rm V} \sim 3.2-7.8$, by assuming the same distance as in this work, \cite{Massi+15_aap573_95} find slopes in the range $\alpha \sim 1.9 - 2.4$ for stellar masses between $\sim 3-6$ and $\sim 60-80$ $M_{\sun}$, depending on the estimator used.

The flatter CMF in the active region would be another signature of the effects of an
intense FUV field, already discussed for the cores in the PDRs in Sect.~\ref{gorti:pdr}.
The FUV field would photodissociate the smaller cores more quickly, resulting in a top-heavy CMF. However, the effect would be more noticeable in the low-mass end of the CMF, which cannot be probed by our observations. This does not necessarily imply that a flatter CMF would in turn result in a flatter IMF as the FUV field may affect the SFE of the cores as well. 

We have assumed isothermal cores. The development of temperature gradients may impact on core mass determination in two ways. Protostellar cores are heated from the inside, thus using a single dust temperature, a mean value weighted by the coldest outermost layers, can lead to overestimating a core mass. More massive cores should be more affected than less massive cores, so a residual protostellar contamination of our starless sample in the quiescent (lower FUV-flux) region may cause the actual CMF to be slightly steeper than found. On the other hand, cores in the active (high FUV-flux) region are heated from the outside, so a single dust temperature is a mean value dominated by the warmer outermost layers, resulting in an underestimate of the mass. A prototellar contamination of the sample would result in cores with both outer and inner heating and would probably just mitigate the effect. Thus the actual CMF in the UV-active region might be even flatter than found. Temperature gradients in cores are therefore likely to enhance the difference between CMFs rather than smooth it.

\subsection{Cloud-cloud collision scenario}

\cite{2018PASJ...70S..41F} (erratum in  \citealt{2018PASJ...70S..60F}) propose that the miniburst of star formation in NGC6334 and NGC6357 originated in a cloud-cloud collision scenario.
It is therefore interesting to check whether the CMFs that we have derived show
any signature of such a process, provided this is still at work. This scenario has been theoretically investigated through
numerical simulations by a number of authors
(e.g. \citealt{2014ApJ...792...63T}, \citealt{2015MNRAS.453.2471B},
\citealt{2018PASJ...70S..58T}, \citealt{2018PASJ...70S..54S}, \citealt{2018PASJ...70S..57W},
\citealt{2021PASJ...73S.405F})
generally finding that cloud-cloud collisions increase the SFE and boost the birth of high-mass stars. 
\cite{2018PASJ...70S..58T} find a CMF with $\alpha = 1.6$ in the high-mass end for a low collision speed ($5$ km s$^{-1}$), which is consistent with our determinations in the active region.  However, when the collision speed is increased, a break will develop on the CMF ($300$ $M_{\sun}$ at
$v = 10$ km s$^{-1}$ shifting to lower masses for higher collision velocities) above which the CMF steepens.
\cite{2018PASJ...70S..54S} derived an IMF steeper than a Salpeter one in the high-mass end (at collision velocities of 10 and 20 km s$^{-1}$).
The inclusion of photoionising feedback by these authors increases the SFE and results in a slightly flatter IMF,
although still steeper than a Salpeter one.

\cite{2018PASJ...70S..41F} find a velocity separation between clouds of $\sim 12$ km s$^{-1}$)
towards NGC6357,
so a comparison between simulations and the 
results summarised in Table~\ref{table:alpha} shows that the CMF in the active region 
is consistent with the cloud-cloud collision scenario simulated by \cite{2018PASJ...70S..58T},
but not with the simulations of \cite{2018PASJ...70S..54S}. 
Nevertheless, the latter
agrees with the high-end IMF derived by \cite{Massi+15_aap573_95} in Pismis~24 using the wider extinction range,
which is actually steeper than a Salpeter one. In addition, \cite{2018PASJ...70S..58T} predict much higher fractions of gas mass in dense cores than we estimated in NGC6357 ($1.4$ \%). 

On the other hand, \cite{2021PASJ...73S.405F} found a top-heavy CMF for
a collision speed $v = 20$ km s$^{-1}$, which appears to be flat up to 6--10 M$_{\sun}$ at $\ge 0.6$ Myr from the beginning of the collision. However, the CMF steepens for $M > 6-10$ M$_{\sun}$. Our derived CMF does not exhibit such a break,
but because of the completeness limit at $\sim 5$ M$_{\sun}$ we cannot be conclusive. \cite{2021PASJ...73S.405F} also propose that a signature of a cloud-cloud collision can be found in the distribution of core-core separation. We constructed this distribution by using the edges of the minimum spanning tree, as in \cite{2021PASJ...73S.405F}. Unfortunately, our spatial resolution does not allow us to resolve their peak at $\sim 0.1$ pc, if any. Our distribution is nearly flat from 0 to $\sim 50$ pc (the complex radius), with a low peak between 10--20 pc that may resemble the second peak yielded by the simulations of \cite{2021PASJ...73S.405F}.
Ultimately, we cannot draw any clear conclusion on a cloud-cloud collision scenario in NGC6357 based only on the comparison between observed and theoretical CMFs.

\section{Summary and conclusions}
\label{sum:conc}

We have used JCMT SCUBA-2 observations at 450 and 850~$\mu$m, complemented with Herschel (Hi-Gal) maps at 70 and 160~$\mu$m, to study the properties of the dense core population in the galactic H\textsc{ii} regions-young star clusters-star forming complex NGC6357. In particular, we aimed to assess the effect of intense FUV radiation on dense cores.

We mapped the region in the \CO(3-2) line as well, to correct the emission at 850 $\mu$m from line contamination in order to construct a map of pure (dust) continuum emission. We then used the algorithm Gaussclumps on the map at 850 $\mu$m to identify the dense cores after removing the extended emission..

We retrieved 1221 cores. For 686 of these we could derive a dust temperature by fitting greybodies to their SEDs obtained from submm (SCUBA-2) and FIR (Herschel) fluxes. The observed (beam-convolved) sizes lie in the range $\sim 13^{\arcsec}$ (the beam size at 850 $\mu$m) to $\sim 30^{\arcsec}$
($\sim 0.1$ to $0.2$ pc); 452 of these have a deconvolved size of at least half the beam size.
We divided the mapped area into two sub-fields, designated as active and quiescent regions, which are roughly the eastern and the western halves of the NGC6357 complex, respectively. The active region contains the three H\textsc{ii} regions G353.2+0.9, G353.2+0.6, and G353.2+0.7 and their associated star clusters. The cores in this region (411) are expected to be affected by an intense FUV field. The quiescent region contains 275 cores farther from the ionising sources, hence less affected by FUV photons. We therefore used the quiescent region as a control sample to study the effects of the FUV radiation on dense cores in the active region. 

Our main results are the following:
\begin{enumerate}
\item We found different core temperature distributions in the two regions, peaking at $\sim 25$ K (quiescent region) and $\sim 35$ K (active region), which we attribute to the effects of the FUV radiation.
\item We further subdivided our sample into starless and protostellar cores, by exploiting the association with protostellar tracers such as WISE point sources and emission at 70 $\mu$m. We found 51 candidate protostellar and 224 starless cores in the quiescent region, and 160 candidate protostellar and 251 starless cores in the active region. Unfortunately, the sample of the active region is biased towards protostellar cores due to the effects of the external FUV field which heats the dust increasing its emission at 70 $\mu$m and saturation in the WISE bands towards the H\textsc{ii} regions. 
\item We derived the core masses by assuming a plausible dust opacity (we also discussed possible effects of FUV radiation). We estimated a mass completeness limit of $\sim 5$ M$_{\sun}$.
We found mass-size relations of
$\log(M/M_{\sun}) \sim a \times \log (D/{\rm arcsec})$, with $a$ in the range $2.0-2.4$, consistent with those from other regions or large samples of clumps. The starless cores above the mass completeness limit are likely to be gravitationally bound, hence pre-stellar in nature.
\item The estimated fraction of molecular gas in dense cores is $1.4$ \% in NGC6357. This is consistent with values from most galactic star formation regions.
\item We showed that the properties of the cores nearer to the H\textsc{ii} regions are consistent with pre-existing cores gradually being engulfed in a PDR and photoevaporating.
\item We constructed the CMFs for $M > 5$ M$_{\sun}$, finding a Salpeter-like CMF in the quiescent region ($\alpha \sim 2.1-2.4$) and a significantly flatter (at a 2 $\sigma$ level) one in the active region ($\sim 1.7-1.9$ with $\alpha = -{\rm d}\log({\rm d}N/{\rm d}M)/{\rm d}\log(M)$). 
The difference becomes even more significant for pre-stellar cores when using the maximum likelihood estimator ($\alpha \sim 2$ vs. $\alpha \sim 1.2$). We attribute this to the effects of the FUV radiation as well. 
\item We compared the CMFs with those predicted by simulations of cloud-cloud collisions, finding no conclusive evidence for cloud-cloud collisions giving rise to the cores, rather than them being pre-existing.
 \end{enumerate}
In conclusion, we found differences in the global properties of the cores in the active region and those in the quiescent region. In particular we found a statistically significant difference in the slope of the CMF between the two regions. We attribute these differences to the influence of the FUV radiation.

\begin{acknowledgements}
The James Clerk Maxwell Telescope was operated by the Joint Astronomy Centre until 2015, on behalf of the Science and Technology Facilities Council of the United Kingdom, the National Research Council of Canada and the Netherlands Organisation for Scientific Research. 
Additional funds for the construction of SCUBA-2 were provided by the Canada Foundation for Innovation. 
The James Clerk Maxwell Telescope has been operated by the East Asian Observatory since 2015, on behalf of The National Astronomical Observatory of Japan, Academia Sinica Institute of Astronomy and Astrophysics, the Korea Astronomy and Space Science Institute, the National Astronomical Observatories of China and the Chinese Academy of Sciences (Grant No. XDB09000000), with additional funding support from the Science and Technology Facilities Council of the United Kingdom and participating universities in the United Kingdom and Canada. 
We thank Harriet Parsons for the Perl script used to correct the 850 $\mu$m data with the CO-emission, and Eugenio Schisano for providing us with the Herschel 160 $\mu$m map used in Fig.~\ref{fig:daisyfields}.
\noindent
This research has made use of NASA's Astrophysics Data System Bibliographic Services (ADS) and of the SIMBAD database, operated at CDS, Strasbourg, France.
\noindent
This work has made use of data from the European Space Agency (ESA)
mission {\it Gaia} (\url{https://www.cosmos.esa.int/gaia}), processed by the {\it Gaia} Data Processing and Analysis Consortium (DPAC,
\url{https://www.cosmos.esa.int/web/gaia/dpac/consortium}). Funding
for the DPAC has been provided by national institutions, in particular
the institutions participating in the {\it Gaia} Multilateral Agreement. AG and FM were partly supported by INAF through the grant Fondi Mainstream 'Heritage of the current revolution in star formation: the Star-forming filamentary Structures in our Galaxy'.
\end{acknowledgements}

\bibliographystyle{aa}
\bibliography{master_biblio.bib}

\newpage

\section{Appendix}

Here we show examples of SED fits to cores in the active and quiescent regions (see Sect.~\ref{pre:co}).

\begin{figure}
\includegraphics[width=\columnwidth]{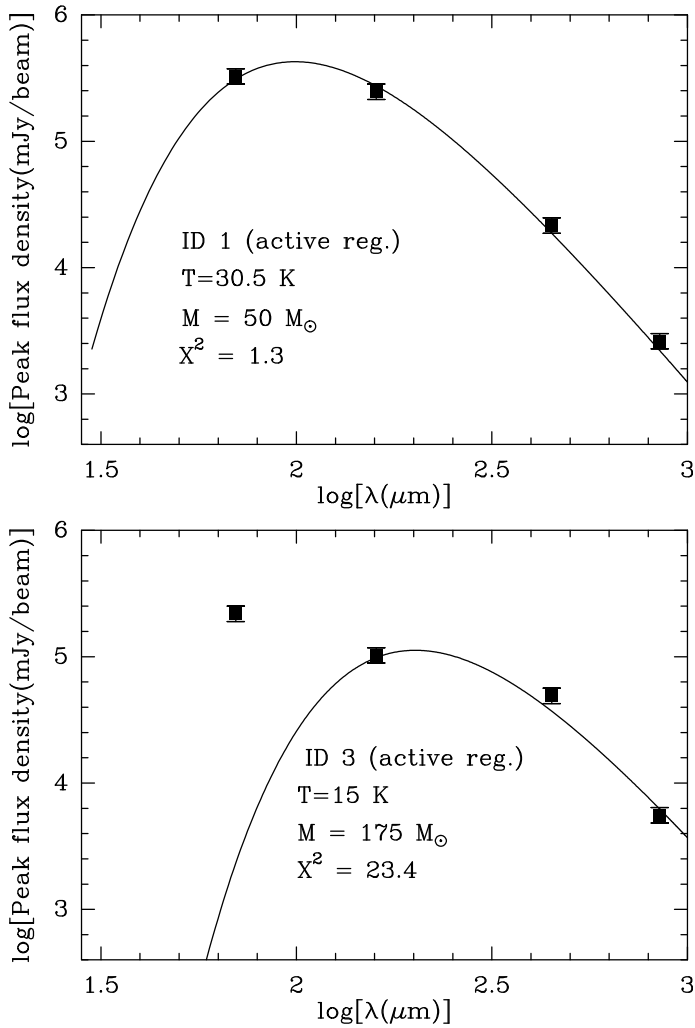}
\caption{Example of SED fits to two cores in the active region. In the top panel, all flux densities 
can be fitted with a single greybody; there is no excess at 70$\mu$m, and this core is 
recognised as a pre-stellar core. The core in the lower panel has an excess flux density at
70$\mu$m (the point has not been excluded from the fit, hence the large $\chi^2$). This could 
be due either to the core being heated from the outside by the FUV radiation field, or to a 
protostar inside the core (see Sect.~\ref{pre:co}).}
\label{fig:SED-A1}
\end{figure}

\begin{figure}
\includegraphics[width=\columnwidth]{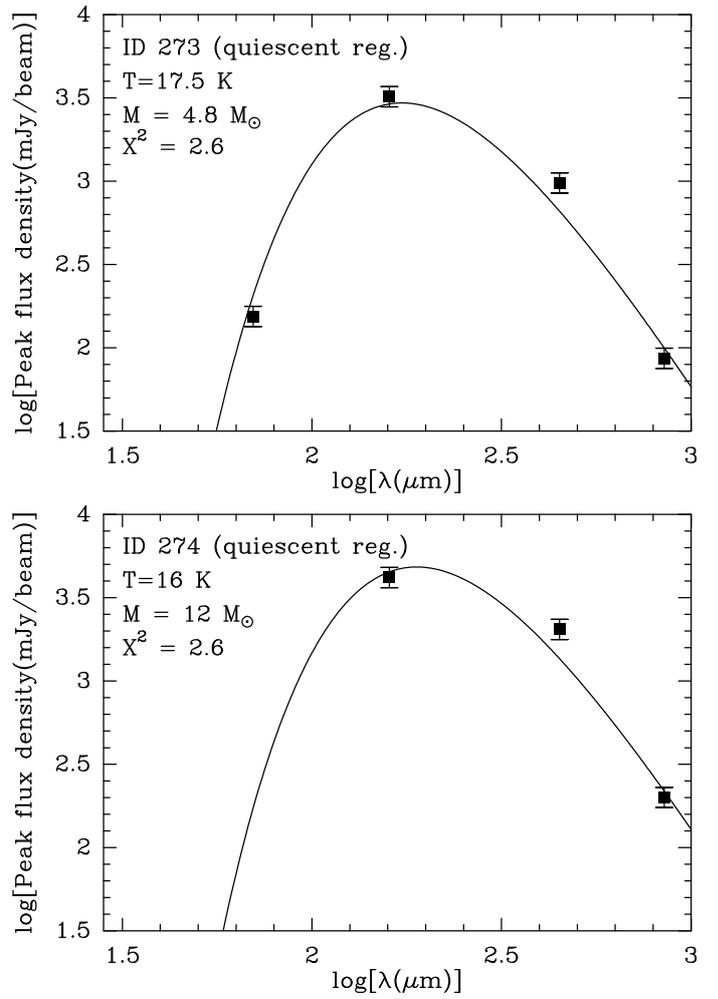}
\caption{Same as Fig.~\ref{fig:SED-A1}, but for two cores in the quiescent region.}
\label{fig:SED-A2}
\end{figure}

\end{document}

%% file: new_commands_aa.tex
\newcommand{\tmb}{T_\mathrm{{MB}}}

\newcommand{\td}{T_\mathrm{{d}}}

\newcommand{\CO}{$^{12}$CO}

\newcommand{\cm}{\usk\centi \metre}

\newcommand{\msun}{\usk\mathrm{M_\odot}}

\newcommand{\jytab}{\mathrm{Jy}}

\newcommand{\kms}{\usk\kilo\metre\usk\second^{-1}}

\newcommand{\kel}{\usk\kelvin}